\documentclass[prb,twocolumn,nofootinbib,longbibliography]{revtex4-2}
\usepackage{graphicx}
\usepackage{lipsum}
\usepackage{multirow}
\usepackage[normalem]{ulem}
\usepackage{braket}
\usepackage{epsfig}
\usepackage{floatrow}
\usepackage{amssymb,amsmath,amsfonts,hyperref}
\usepackage{wasysym}
\usepackage{latexsym}
\usepackage{xcolor}
\definecolor{je}{rgb}{0.858, 0.188, 0.478}
\definecolor{so}{rgb}{0.19,0.32,1.52}
\usepackage{verbatim}

\usepackage[caption=false]{subfig}
\usepackage{braket}
\usepackage{lipsum}
\usepackage{cleveref}

\def\Tr{\mathrm{Tr}}
\newcommand{\xb}{\bar{x}}
\newcommand{\yb}{\bar{y}}
\newcommand{\ff}{\mathbb{F}_2}
\newcommand{\dm}{\mathcal{D}}
\newcommand{\dmb}{\tilde{\mathcal{D}}}
\newcommand{\zz}{\mathbb{Z}}
\newcommand{\tn}{2^{n}}
\newfloatcommand{capbtabbox}{table}[][\FBwidth]
\newcommand{\ts}[1]{\textsubscript{#1}}   
\newcommand{\ssl}[1]{s_{\alpha^{#1}}}

\begin{document}

\title{Beyond the Freshman's Dream: Classical fractal spin  liquids from matrix cellular automata in three-dimensional lattice models}
\author{Sounak Biswas}\thanks{These authors contributed equally.} 
\affiliation{Rudolf Peierls  Centre  for  Theoretical  Physics, Parks Road, Oxford  OX1  3PU,  United  Kingdom}
\author{Yves H. Kwan}\thanks{These authors contributed equally.} 
\affiliation{Rudolf Peierls  Centre  for  Theoretical  Physics, Parks Road, Oxford  OX1  3PU,  United  Kingdom}
\author{S. A. Parameswaran}
\affiliation{Rudolf Peierls Centre for Theoretical Physics, Parks Road, Oxford OX1 3PU, United Kingdom}

\begin{abstract}
  We construct models hosting classical fractal spin liquids on two realistic three-dimensional (3D) lattices of corner-sharing triangles: trillium and hyperhyperkagome (HHK).   Both models involve the same  form of three-spin Ising interactions on  triangular plaquettes as the Newman-Moore (NM) model on the 2D triangular lattice. However, in contrast to the NM model and its 3D generalizations, their  degenerate ground states and low-lying excitations {\it cannot} be described in terms of scalar cellular automata (CA), because the corresponding fractal structures lack a simplifying algebraic property, often termed the `Freshman's dream'. 
By identifying a link to  \textit{matrix} CAs --- that makes essential use of the crystallographic structure --- we show that both models exhibit fractal symmetries of a distinct class to  the NM-type models. We devise a procedure to explicitly construct low-energy excitations consisting of finite sets of  immobile defects or ``fractons",  by flipping  arbitrarily large self-similar subsets of spins, whose fractal dimensions we compute analytically. 
 We show that  these excitations are associated with energetic barriers which increase logarithmically with system size, leading to ``fragile'' glassy dynamics, whose existence we confirm via classical Monte Carlo simulations. We also discuss consequences for spontaneous fractal symmetry breaking when quantum fluctuations are introduced by a transverse magnetic field, and propose multi-spin correlation function diagnostics for such transitions. Our findings suggest that  matrix CAs may provide a fruitful route to identifying fractal symmetries and fracton-like behaviour in  lattice models, with possible implications for the study of  fracton topological order.
\end{abstract}
\maketitle

\section{Introduction}

Fractals and other self-similar structures are ubiquitous in  systems of many interacting constituents, appearing in contexts as diverse as localisation physics, percolation theory, complex networks, and in the study of phase transitions~\cite{Mandelbrot,BundiEA,Castellani,StaufferAharony}. In the condensed matter setting,  self-similarity is usually a consequence of scale invariance, most often realised by tuning the system to a  critical point or manifold in parameter space~\cite{GrossmanAharony,StellaVanderzande,DuplantierSaleur,Duplantier,JankeSchakel}.  Since continuous transitions are associated with a diverging  correlation length, the physics proximate to them usually admits a coarse-grained description in terms of continuum fields. The continuum theory typically has more symmetry than the underlying microscopic model,  for instance exhibiting {\it continuous} rotational, translational, and scale invariance even when the microscopic degrees of freedom live on a discrete lattice. Fractality is  usually  only manifest only in this  continuum scaling limit, and it is generally difficult to directly describe the emergent self-similar structures in terms of the original microscopic lattice degrees of freedom. 

In this paper, we identify and study two  three-dimensional lattice models that are exceptions to this general picture, in that their ground states and low-lying excitations exhibit fractal properties that are manifest directly at the  lattice scale. This is linked to the fact that the models enjoy a set of exact \textit{fractal symmetries} whose  transformations act on a fractal subset of the degrees of freedom~\cite{Yoshida,TrithepEA}. At low temperatures, these models can be viewed as ``classical fractal spin liquids'' with a macroscopic, but subextensive, number of  classical ground states,  each with fractal sets of flipped spins relative to the uniform ferromagnetic configuration.  We demonstrate that these properties can be understood in terms of matrix cellular automata, where the matrix indices are associated with the basis of the underlying (non-Bravais) crystal lattice. Our analysis lays the groundwork for a  systematic search for similar properties in other realistic lattice systems. 

Our work  builds significantly on a  two-dimensional classical statistical mechanical model with similar properties, first studied by Newman and Moore (NM)~\cite{Newman_Moore,Garrahan_Newman}. The NM model describes Ising spins placed on the sites of a  triangular lattice and coupled by three-spin interactions only on the up-pointing triangular plaquettes; an identical model but with interactions on all plaquettes was  famously studied by Baxter and Wu (BW)~\cite{BW_PRL,BW_AJP1} and displays a second-order phase transition in the universality class of the 4-state Potts model. However, unlike the BW model, the NM model can be mapped to a free gas of ``defect spins'' associated with energetically unfavorable plaquettes, and hence has trivial thermodynamics. Despite this, endowing the NM model with classical single-spin-flip dynamics leads to glassy behaviour  even in the absence of disorder.  The glassiness stems from the fractal structure of the low-lying excitations, which makes the
 activation barriers to equilibration grow logarithmically, and hence equilibration timescales as a power law, with system size. Such equilibration times have an exponential inverse-temperature square (EITS) dependence $\tau \sim \exp(1/T^2)$. This super-Arrhenius behaviour, sometimes referred to as ``fragile" glassiness, indicates that activation barrier heights increase with decreasing temperatures~\cite{Angell_Science,Debenedetti_Nature}. The dynamics is hierarchical, with fast degrees of freedom activating slow degrees of freedom, much like  kinetically constrained models of glass-formers~\cite{KCM_review} such as the East model; however, in the NM model the constraints on the microscopic dynamics are not imposed explicitly but rather emerge from the local energetics. As an exactly-solvable model which leads to glassiness from local constraints, the NM model has been the subject of several investigations into its  properties and generalizations ~\cite{Turner_Jack_Garrahan,Jack_Garrahan,Garrahan,BiroliEA}. Since they lack low-temperature order yet exhibit fractal correlations as $T\to 0$,  cooperative paramagnets with such phenomenology have been dubbed ``classical fractal spin liquids''.

Previous attempts to generalize the NM model to 3D have considered  complex higher-spin interactions on  lattices designed specifically to admit an analysis in terms of linear {\it scalar} cellular automata (CA), a standard route to generating fractal structures~\cite{Willson}. In the context of lattice spin models such as NM, the CA prescribes an iterative rule for constructing ground states or low-lying excitations by flipping spins in a layer of a lattice given those flipped in the preceding layer.  The scalar CA describing the NM model satisfies a mathematical identity known as the ``Freshman's dream''. This property  --- so named for its appealingly simple mathematical expression [Eq. \eqref{eq:freshmans_dream} below] when the action of the  CA is represented by multiplying polynomials over the finite field $\ff$ --- encodes the fact that when initialized with a single flipped spin, iterating the CA $2^n$ times leads to a configuration with only a finite number of  plaquettes excited out of their ground state. The simplifications enabled by the Freshman's dream combined with linearity of the CA allow an  essentially complete analysis of fractonic excitations in the NM model, and the attendant consequences for both thermodynamics and dynamics.

The distinction between the existing 3D generalizations of the NM model and those studied in this paper lies in the nature of the CA that characterizes the models. Rather than deform the model and lattice to ensure a description in terms of scalar CAs, we instead identify two different lattices of corner-sharing triangles ---  trillium~\cite{SchroderEA,PfleidererEA}, and hyperhyperkagome (HHK) ~\cite{NakamuraEA,KhuntiaEA,ChillalEA} --- on which the Baxter-Wu three-spin interaction can be imposed on each plaquette. Both of these lattices have been previously investigated with Heisenberg-type nearest-neighbor spin-spin interactions, in the context of seeking classical and quantum spin liquids stabilized by geometrical frustration~\cite{Hopkinson_Kee,Isakov_Hopkinson_Kee,Canals_Lacroix,ChillalEA,KhuntiaEA,Chern_Kim,Jin_Zhou}. We find  that rather than the `conventional' classical spin liquid behaviour with an extensive $T=0$ entropy, the trillium and HHK Baxter-Wu models instead exhibit fractal symmetries with  ground state degeneracies that contribute subextensively to the entropy as $T\to 0$, a hallmark of a classical fractal spin liquid of the NM type. However, the appealing simplicity of the models and lattices comes at a price: we find that the relevant fractals are now described by matrix CAs, i.e., the transition function describing the CA is a matrix, and the transition rule does not satisfy the Freshman's dream. We nevertheless show that we may use the algebraic properties of these transition matrices to construct  immobile excitations with fractal structure. We also deploy  the matrix CA technology  to demonstrate both the triviality of the thermodynamics and the glassiness of the dynamics, as well as to construct a variety of correlation-function measures of symmetry breaking. Thus, a modest increase in the complexity of the CA description (and associated changes in the nature of the fractal operators) allows us to develop a ``lattice-first'' approach that admits simpler interactions and applies to more realistic crystal structures than have been previously explored in this context.
 
As we have noted, we study low-energy excitations consisting of sets of immobile defects created by acting on a ground state configuration by a fractal operator. In a different and more quantum-mechanical context, similar excitations have been termed ``Type II fractons''~\cite{Nandkishore_Hermele,Pretko_Chen_You}, and  have been the  subject of intensive study  in recent years
~\cite{Chamon,BravyiEA,Vijay_Haah_Fu_X3,Haah,Yoshida,Vijay_Haah_Fu, Pretko1,Pretko2,Ma_Hermele_Chen,Bulmash_Barkeshli,Bulmash_Barkeshli_fractal,Fontana_Gomes_Chamon}. 
 Like the low-energy excitations of the NM model and its cousins studied in this paper, type II fracton phases  have point-like defects lying at the corners of fractal operators~\cite{Haah,Yoshida}, and exhibit super-Arrhenius relaxation and glassiness due to the presence of logarithmic barriers~\cite{Castelnovo_Chamon,Prem_Haah_Nandkishore}. However, unlike their classical counterparts, models hosting type II fracton phases are also topologically ordered: their degenerate ground space on the torus includes states unrelated by symmetries, leading to a topological  degeneracy protected against arbitrary local perturbations. Although the classical models we study here lack this topological structure, they nevertheless host  immobile defects linked to fractal operators; so, in a mild abuse of terminology, we will refer to such defects as fractons throughout this paper.

Quantum fluctuations can be introduced into a classical fractal spin liquid  applying a transverse magnetic field~\cite{YizhiEA, Yoshida_Kubica_fractal}. As in the case of the Ising model in one spatial dimension, this allows us to drive a genuine $T=0$ quantum phase transition despite the trivial thermodynamics for $T>0$.  At large transverse field, the system realises a trivial paramagnet, but upon decreasing the  field it  undergoes a fractal quantum phase transition between phases where the fractal symmetries are preserved and one where they are broken. Recent work revisiting this scenario in the 2D NM model has suggested  that the transition between these phases is continuous, with unusual scaling behaviour near criticality rationalized in terms of ``UV-IR mixing'' linked to the fracton excitations~\cite{YizhiEA}. While we defer a full numerical exploration of such a transition in our 3D models to future work, we lay the foundations for such a study by identifying the appropriate multispin correlation function diagnostics for fractal symmetry breaking in our lattice models. In the scenario just outlined, the high-field phase is a quantum paramagnet, while the low-field phase spontaneously breaks fractal symmetry. On some lattice models with fractal subsystem symmetries, it is possible to implement a so-called ``F-S duality'' transformation~\cite{Vijay_Haah_Fu}, which yields a quantum Hamiltonian made up of commuting stabilizers that, rather than paramagnetic or broken-symmetry order, instead exhibits Type II fracton topological order. However, as we explain below,  neither of the two models introduced in this paper is F-S dual to a fracton phase.

The remainder of this paper is organized as follows. We begin with a review of NM model, its fractal symmetries, thermodynamics, and glassy classical dynamics in  Sec.~\ref{sec:review}, focusing on how these can be understood from the lens of scalar CA. In Sec.~\ref{sec:crystal_structure}, we introduce the models under investigation in this paper on two different lattices of corner-sharing triangles: trillium and HHK. In Sec.~\ref{sec:CA}, we show that these models are fractal symmetric, and derive the matrix CA which describe their ground states. In Sec.~\ref{sec:fractons}, we use these matrix CA to explicitly construct fracton defects, and show that the energy barriers associated with creating such fractons scale logarithmically with system size. In Sec.~\ref{subsec:thermodynamics}, we solve for the thermodynamics of these models. In Sec.~\ref{subsec:glassiness} we estimate the heights of barriers to equilibration, which grow logarithmically with system size, and numerically show that these barriers result in glassy classical dynamics. In Sec.~\ref{subsec:multispin_corrs}, we consider implications for fractal symmetry-breaking in quantum generalisations of our models, and propose multi-spin correlation functions to diagnose such phase transitions. We close with a summary and a survey of future directions.

\section{The Newman-Moore model, cellular automata, and fractons}
\label{sec:review}

This section is intended to serve two purposes---the first is to review fractal symmetries, fractonic excitations, and glassiness in the familiar NM model;   the second is to introduce linear CA in the language of polynomials over the finite field $\ff$, which proves  to be an indispensable tool in our investigations of BW models on trillium and HHK. We draw liberally from the discussions in Refs.~\cite{Newman_Moore, Yoshida,TrithepEA}; readers familiar with these papers and with the CA technology can skip this section, but may nevertheless find it useful to skim it to remind themselves of key results or to acquaint themselves with our notation and conventions.

The NM model describes Ising spins placed on the sites of a 2D triangular lattice, with three-spin interactions on all \emph{up}-pointing triangular plaquettes, described by the classical Hamiltonian
\begin{align}
  H =-\sum_{(i,j)}  \sigma(i,j) \sigma(i-1,j-1) \sigma(i,j-1).\label{eq:NM-Ham}
\end{align}
Here, $\sigma(i,j) =\pm 1$ is an Ising spin at a lattice site $(i,j)$, which is placed at spatial location $\mathbf{r}_{(i,j)} = i \mathbf{a}_1 + j \mathbf{a}_2$, where $\mathbf{a}_1 = \hat{\mathbf{x}}$ and $\mathbf{a}_2 = -\frac{1}{2}\hat{\mathbf{x}}+\frac{\sqrt{3}}{2}\hat{\mathbf{y}}$ are primitive vectors of the triangular lattice. 
Note that this coordinate system is chosen so that the spins $\sigma(\cdot,j)$ live on the row \emph{above} $\sigma(\cdot,j-1)$. 
It will be convenient to also define binary variables $s(i,j)$ which take values in $\{0,1\}$. 

The emergence of fractal symmetries in the NM model  stems from the fact that constraining a configuration to be a ground state of \eqref{eq:NM-Ham} uniquely specifies $s(i,j)$ in terms of $s(i,j-1)$ and $s(i-1,j-1)$ on the row below. Fixing all spins at a particular value of $j=p$ therefore fixes all spins at values of $j>p$, i.e. above it. The ground state degeneracy corresponds to the freedom of choosing the initial layer of spins at $j=0$. The exact number of ground states depend on the exact system size and boundary conditions for finite systems, as outlined later. [We use the term `ground state' to refer to any classical configuration  that minimizes~\eqref{eq:NM-Ham}.]

The configuration described by $s(i,j)=0$ [i.e. $\sigma(i,j)=1$] for all $(i,j)$ is trivially a ground state, and we will often refer to $s(i,j)=1$ as a `down' or `flipped' spin at the location $(i,j)$.
It is convenient to use the language of polynomials over the finite field $\ff$, which we briefly introduce here, in preparation for its extensive use throughout the rest of the paper.  
A spin configuration $s(i,j)$ can described by the set of polynomials $u_j (x)=\sum_i s(i,j) x^i$, where $s(i,j) \in \ff$. Observe that a translation along the $\hat{\mathbf{x}}$ direction is represented by multiplication by $x$, so that each polynomial $u_j$  describes the spins along a fixed $j$-slice. While any lattice spin system can be given such a labelling, a special feature of NM and related models is that the different $u_j$'s are not independent for a ground state. Instead, the ground state constraint determines a layer of spins from the previous layer, so that successive configurations $u_j$ can be described by iterating a linear CA that describes each spin $s(i,j)$ in terms of a few spins  $s(i',j-1)$ in the previous layer, where $i'$ is in some local neighbourhood of $i$.  
In terms of polynomials over $\ff$, the CA that generates ground state configurations of the NM model is given by
 \begin{equation}
u_{j+1} (x)=f(x) u_{j}(x), \end{equation}
with the \textit{transition function}
 \begin{equation}
f(x)=1+x.
\end{equation}
Each distinct ground state is associated to a fractal symmetry of the system, implemented by flipping spins at the positions of its down spins, i.e.  the  operation that generates the ground state from the ``all up'' configuration.

If the system is defined on a finite cylinder with periodic dimension $L_x$, the ground state degeneracy is $2^{L_x}$, corresponding to the free choice of initial spins on layer $j=0$, since all the other spins are uniquely fixed once these are specified.

On a torus, the exact number of ground states is given by the possible solutions $u(x)$ to the equation
\begin{align}
\nonumber  f(x)^{L_y} u(x)&= u(x),\\
  x^{L_x}&=1.
  \label{eq:gs_cond}
\end{align}
In particular, for $L_x=L_y=2^n$ (more generally $L_y\geq L_x=2^n$), the only solution is $u(x)=0$, which corresponds to $s(i,j)=0$ for all $(i,j)$ i.e. the `all up'' configuration. This result follows from the Freshman's dream, a key property of polynomials over $\ff$, defined via the identity
\begin{align}
  (\sum_i a_i x^i)^{2^n} =\sum_i a_i x^{i 2^n},  n \in \zz.
  \label{eq:freshmans_dream}
\end{align}

Since $x^0$ and $x^{2^n}$ are identified on the torus when $L_x =2^n$, Eq.~\ref{eq:freshmans_dream} gives $f(x)^{L_y}=0$ and hence $u(x)=0$ is the only solution of~\eqref{eq:gs_cond}. 

We introduce defect variables by 
\begin{equation}\tau=(s(i,j)+s(i,j-1)+s(i-1,j-1)) \mod 2,\end{equation}
 corresponding to plaquettes that violate the ground state condition. A ground state is thus specified by the absence of defects. Since the number of  defect variables (defined on up-pointing triangles) is equal to the number of spins, the existence of a unique ground state implies a bijection between spin and defect configurations.\footnote{If there are $W$ distinct ground states, then the spin-to-defect mapping is $W$-to-1.} The defect variables are non-interacting,  with trivial thermodynamics controlled by the free-defect partition function $Z=(\cosh(\beta J))^{L_x L_y}$. While the solution is exact for tori of size $L_x=L_y=2^n$, the deviations for other system sizes are sub-extensive and consequently the thermodynamic limit is trivial. Note, however, that under this mapping the single-spin flip dynamics create triplets of defects; conversely, a {\it single} defect cannot be relaxed by the action of a local operator. Consequently, under single-spin flip dynamics (assumed for the rest of this paper), a state with a finite density of isolated defects will relax very slowly to equilibrium, i.e. the dynamics are glassy.

A more quantitative understanding of the relaxation can be obtained by analyzing the self-similar fractal patterns generated by the CA. Consider a  CA described by a transition function $f(x)=\sum a_i x^i$, with $N$ nonzero values of $a_i$. If one starts with $u_0(x)=1$ (a single down spin at the origin), it follows from Eq.~\eqref{eq:freshmans_dream}  that $u_{2^n}(x)=\sum_i a_i x^{i2^n}$. Therefore rows $j=2^n$ host  exactly $N$ down spins, corresponding to the nonzero values of $a_i$, but rows with $j\neq 2^n$ will generically host $O(j)$ down spins. Further CA evolutions from these flipped spins in layer $j=2^n$ resemble the initial evolution from the point $(i,j)=(0,0)$, resulting in a self-similar structure. For the NM transition function $f(x)=1+x$, these further evolutions do not overlap until $j=2^{n+1}$, i.e. the evolution at depth $n$ terminates before that at depth $n+1$ begins,  leading to a simple fractal known as the Sierpinski triangle. This `nonoverlapping' nature of the self-similar structure is one reason why the NM model is so simple.   

\begin{figure}
    \includegraphics[width=\textwidth]{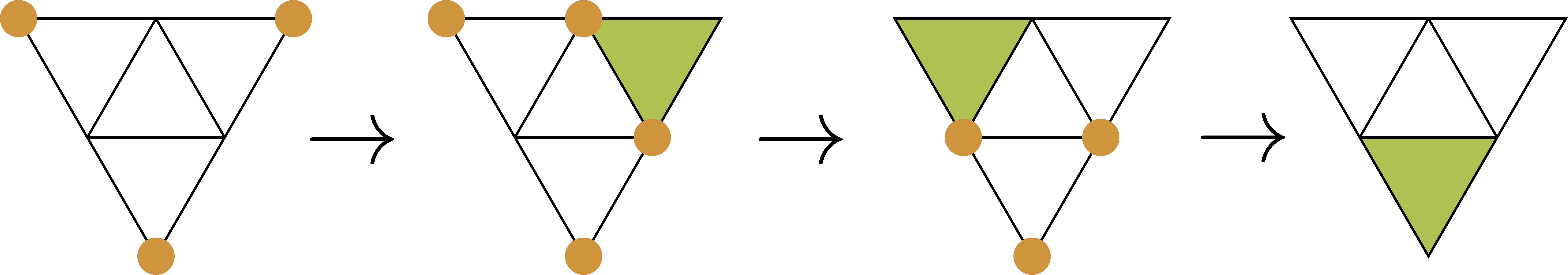}
    \caption{Annealing a $2^n$-sized three-fracton excitation (left) away by sequentially applying spin-flips which create $2^{n-1}$-sized excitations requires an intermediate $4$-defect configuration. Yellow dots indicate fractons. Green shading represents the application of a Sierpinski triangle of spin-flips.}
    \label{fig:NM_paper}
\end{figure}

The fractal structure  implies the existence of ``fractonic" excitations which, when combined with the properties of  single-spin-flip dynamics,  leads to glassy behaviour. Consider a starting configuration with a single down spin: $s(0,0)=1$, with all other spins on rows $j\leq0$ being $s(i,j)=0$,  corresponding to the presence of a  single defect on the plaquette below $(0,0)$, with corners  at $[(0,0), (-1,-1), (0,-1)]$. In the polynomial representation, we have $u_0(x)=1$, $u_{j<0}(x)=0$. Now,  fix the spins for each row $j>0$ by using the CA

until $j=2^n$, for some $n \in \zz$. The CA evolution guarantees that no new defects are created in the plaquettes between $j=0$ and $j=2^n$. At $j=2^n$ we have $u_{2^n}(x)=1+x^{2^n}$, which describes 2 down spins at $i=0$ and $i=\tn$. Re-flipping these two spins  back to $0$ creates exactly one defect in each of the two up-pointing triangles below these spins. If we also set $s(i,j)=0$ for all $j>2^n$, we have an excitation with three defects,  and size $2^n$, created by flipping a $2^n$-sized Sierpinski triangle of spins. Note  that we cannot stop the evolution in this manner for generic $j$  without incurring an energy cost that scales with $j$, since there will be generically $O(j)$ flipped spins unless $j= 2^n$ for some $n$. Therefore, under single spin flip dynamics, relaxing a $3$-defect configuration from  `depth' $n+1$ to depth $n$ always involves crossing an intermediate configuration at a higher energy, slowing down the dynamics. As a consequence of this energy landscape, an individual defect or `fracton' is not mobile at all, and must decay into multiple fractons to move. Excitations which are composed of multiple fractons are also immobile.

We can determine the energy barriers for relaxing the 3-fracton configurations from their self-similar structure. Observe that a 3-fracton configuration at depth $n+1$ can be viewed as comprised of a superposition of three 3-fracton configurations at depth  $n$, since these overlap precisely at the internal corners.  For the NM model, one can demonstrate that the minimum-energy path for flipping the depth-$(n+1)$ configuration involves sequentially flipping its three constituent depth-$n$ configurations; however, in the process of doing so, one always encounters an intermediate configuration with 4 fractons, i.e. at energy $J=1$ relative to the initial and final configurations. Consequently, the activation barrier encountered at depth $n+1$ is $1$ higher than that at depth $n$, viz.
\begin{equation}
E_A(n+1) = E_A(n)+1.
\end{equation}
leading to $E_A(n) = n$. Now, recalling that the linear size of a 3-fracton configuration at depth $n$ is $\ell = 2^n$, we see that the energy barrier to relax an excitation of linear size $\ell$ scales as $\log_2 \ell$.

Standard arguments~\cite{Garrahan_Newman,Castelnovo_Chamon,Prem_Haah_Nandkishore}, that we review in Sec.~\ref{subsec:glassiness}, link this scaling with an exponential inverse-square behaviour for equilibration times $\tau \sim \exp(1/[2T^2 \ln 2])$, a phenomenology termed ``fragile glassiness''.

All the above statements generalise in a straightforward manner to the square-pyramid model, which has $5$-spin interactions on the up-pointing square-based pyramids of a BCC lattice~\cite{Turner_Jack_Garrahan}. The lowest energy excitations have  $5$ fractons , and they live on the corners of a fractal of flipped spins in the shape of a Sierpinski pyramid, such that each pyramid is built out of $5$ smaller non-overlapping pyramids of half its size.   Note that this generalizes the `Sierpinskian' fractal structure of the NM model to three dimensions, but at the cost of a substantially more complex interaction. In the balance of this paper, we will pursue a different strategy,  keeping the three-spin interactions of the NM model but exploring different lattices with the corner-sharing-triangle structure of NM, but in 3D.

While the models explored in this paper share many properties with the NM model and its generalizations, they differ in some ways owing to the more complex structure of their CA description. Most notably, the matrix CAs describing the trillium and HHK models  do not  satisfy `Freshman's dream', which considerably complicates the extraction of the properties of ground states with periodic boundary conditions, and of the fractal structure of defect configurations. Another important difference lies in the computation of energy barriers.
 A crucial aspect that enabled their exact computation in the NM model is that 
  when sequentially flipping 3 depth-$n$ configurations in order to relax a depth-$(n+1)$ configuration, each of the intermediate depth-$n$ flips involves a barrier no greater than that experienced by the first, so that the only increase in the barrier energy comes from the 4-defect intermediate configuration. As we will argue, this is not the case in the more complex models we explore here, and consequently we cannot perform a similar  computation in those cases. (Although we nevertheless  extract the effective energy barriers by a different approach.)

\section{Models and lattice structures}
\label{sec:crystal_structure}
We will consider Ising models with three spin interactions, $H=-\sum_{\Delta_{ijk}} \sigma_i \sigma_j \sigma_k$, which couple spins on all triangular plaquettes with vertices $(i,j,k)$, on  two three-dimensional lattices of corner-sharing triangles, known as ``trillium'' and hyper-hyperkagome (HHK). When triads of neighbouring (neighbouring and next-neighbouring) sites are grouped into triangular plaquettes in trillium (HHK), both lattices involve corner sharing triangles such that each spin is a part of three different triangular plaquettes. We now proceed to describe the crystal structures of these lattices.

\subsection{Trillium}
\label{subsec:crystal_trillium}
\begin{figure}[t]
  \includegraphics[width=0.7\columnwidth]{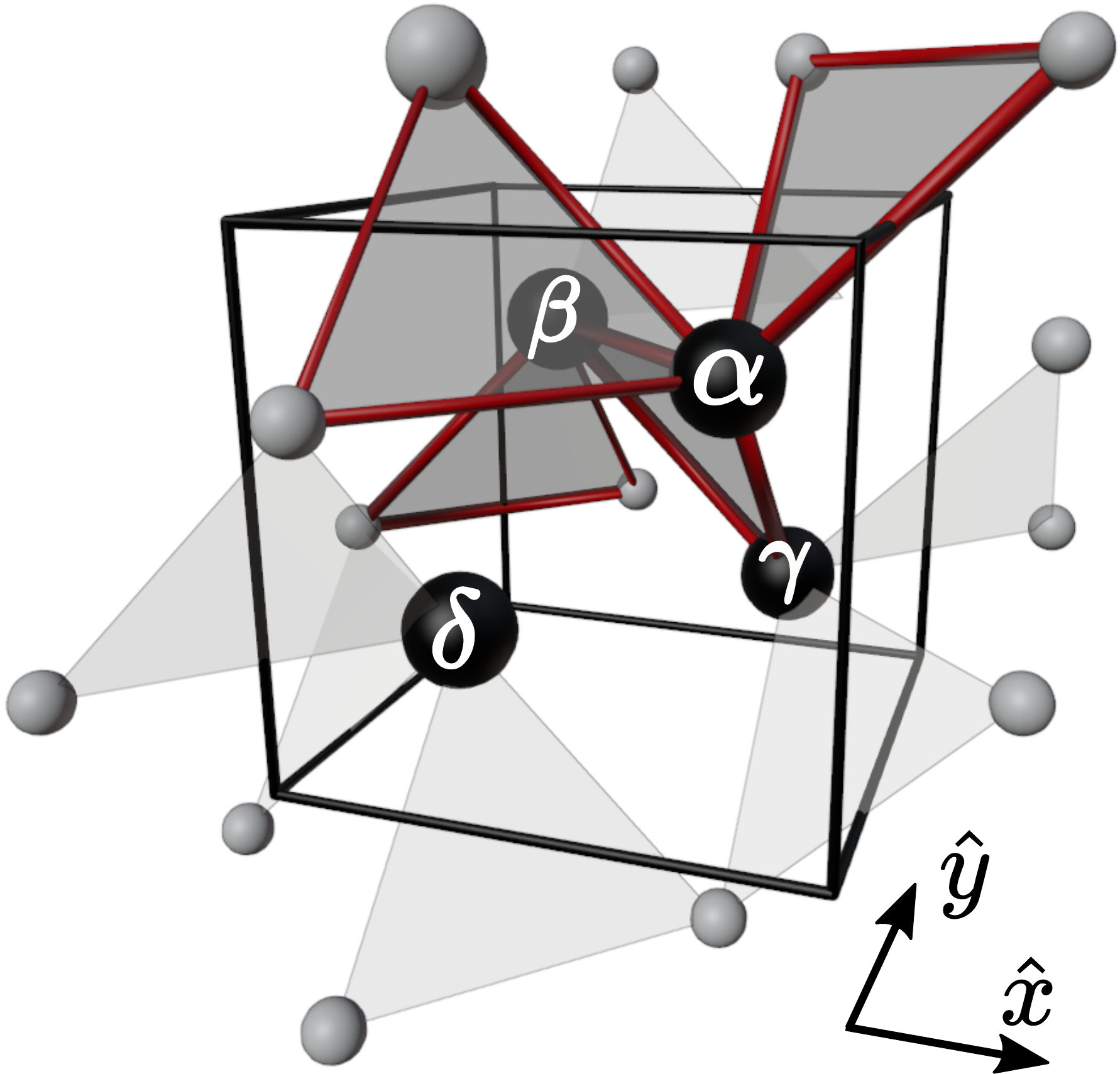}
\caption{\label{fig:trillium}The lattice structure of trillium, with the four sublattices in the home unit cell labelled by $\alpha,\beta,\gamma,\delta$. Grey balls indicate sites outside the home cell. One plaquette of each type is outlined in red.}
\end{figure}

Trillium is a substructure of many systems with simple cubic symmetry group $P2_13$ (No. 198), as exemplified for instance by the Mn ions in MnSi. The sites occupy the 4a Wyckoff position, which has a single free parameter $u$. Each site has $6$ nearest neighbours. If triads of nearest neighbours are grouped into triangular plaquettes, 

then each site belongs of three plaquettes, as in the NM model.  A convenient choice for visualization is to pick $u=\frac{1}{4}$, since the sites then lie at regular positions. The crystal structure is shown in Fig.~\ref{fig:trillium}.  

Given a lattice of corner-sharing triangles, we can construct its dual as follows. 
Each  triangular plaquette is associated with a dual lattice site at its centre.  and dual lattice links are associated with a site on the direct lattice shared by neighboring plaquettes on the dual lattice. Under this duality mapping, the trillium lattice is self-dual; this can be verified by explicit construction as we now briefly summarize.

We follow the labelling conventions of Hopkinson and Kee (Ref.~\onlinecite{Hopkinson_Kee}), and label  the four sublattices  $\alpha,\beta,\gamma$ and $\delta$. We can identify four different types of triangular plaquettes on trillium:

\begin{align}
A_{0,0,0}={}&\alpha_{0,0,0},\beta_{0,0,0},\gamma_{0,0,0}\nonumber\\
B_{0,0,0}={}&\delta_{0,0,1},\beta_{0,-1,0},\alpha_{0,0,0}\nonumber\\
C_{0,0,0}={}&\gamma_{0,0,1},\delta_{1,0,1},\alpha_{0,0,0}\nonumber\\
D_{0,0,0}={}&\beta_{0,0,1},\gamma_{-1,0,1},\delta_{0,1,1}.\label{eq:plaquettes}
\end{align} 
Tab.~\ref{tab:trillium} lists the three plaquettes that are incident to the sites from each of the four sublattices in the primitive unit cell at $(0,0,0)$. Comparing this to the identification of plaquettes in \eqref{eq:plaquettes}, the lattice is evidently self-dual under the mapping $\alpha\leftrightarrow A$, $\beta\leftrightarrow B$, $\gamma\leftrightarrow C$, $\delta\leftrightarrow D$ and a $C_2$ rotation about the $x$-axis, under which $(x,y,z) \leftrightarrow (x, -y,-z)$.

\begin{table}
  \centering
      \begin{tabular}{|c||c|c|c|}
        \hline
	\multicolumn{4}{|c|}{Plaquettes in trillium} \\
        \hline 
	Site  & Plaquette 1 &Plaquette 2 &Plaquette 3  \\ 
	\hline\hline
	$\alpha$                      & $A_{0,0,0}$ & $B_{0,0,0}$ & $C_{0,0,0}$ \\
        \hline
	$\beta$                       & $D_{0,0,-1}$ & $B_{0,1,0}$ & $A_{0,0,0}$\\
        \hline
	$\gamma$                       & $C_{0,0,-1}$ & $D_{1,0,-1}$ & $A_{0,0,0}$            \\
        \hline
	$\delta$                       & $B_{0,0,-1}$ & $C_{-1,0,-1}$ &  $D_{0,-1,-1}$                \\
        \hline
	\end{tabular}
\caption{List of plaquettes (defined in \eqref{eq:plaquettes}) connected to the sites of all four sublattices with unit cell coordinates $(0,0,0)$ in the trillium network of corner-sharing triangles. Subscripts denote unit cell coordinates.}
\label{tab:trillium}
\end{table}

\subsection{HHK}
\begin{figure}
    \includegraphics[width=0.5\textwidth]{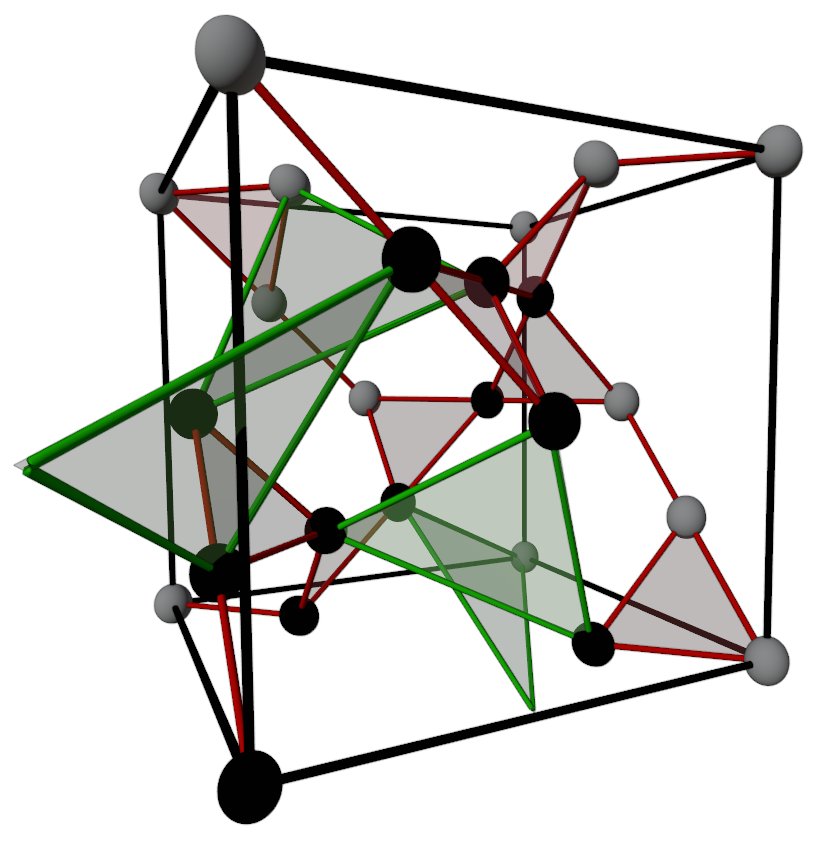}
    
    \caption{\label{fig:hhk}The cubic unit cell of HHK network of corner-sharing triangular plaquettes, with 12 sites in a unit cell. Each site is shared by three triangular plaquettes. }
\end{figure}

The second lattice we consider is the hyperhyperkagome (HHK) or `distorted-windmill' structure. This is realized by the magnetic lattice of $\beta$-Mn as well as $\mathrm{PbCuTe}_{2}\mathrm{O}_6$, both of which have been investigated in the context of frustrated magnetism and spin liquids~\cite{Isakov_Hopkinson_Kee,ChillalEA,KhuntiaEA,Chern_Kim,Jin_Zhou}. The structure corresponds to a simple cubic symmetry group $P{4_123}$, with the sites occupying the special $12d$-Wyckoff positions. If triads of nearest and next-nearest neighbours are grouped into triangular plaquettes, we obtain the HHK structure of corner-sharing triangular plaquettes. Each site resides at the shared corner of three plaquettes. We label the $12$ sublattices $\alpha^i$, for $i \in (0\cdots 11)$. The connections are shown in Tab.~\ref{tab:hhk}. [Note that one can also view HHK as a decoration of the closely-related hyperkagome structure of corner-sharing triangles, but where each site is shared by two rather than three triangles.]

The crystal structure is displayed in Fig.~\ref{fig:hhk}, along with the positions of each sublattice $\alpha^i$. The connectivity, along with the grouping of nearest and next-nearest neighbours into plaquettes, is shown in Tab.~\ref{tab:hhk}.
In contrast to the trillium lattice, this network of corner-sharing triangles is not self-dual, as demonstrated in Appendix~\ref{app:hhk_no_self_duality}. 

\begin{table}
\begin{tabular}{ |c||c|c|c|  }
	\hline
	\multicolumn{4}{|c|}{Plaquettes in HHK} \\
	\hline
	Site & Plaquette 1 & Plaquette 2 & Plaquette 3 \\
	\hline \hline
	$\alpha^0$ & $\alpha^9$\ts{-100} $\alpha^1$\ts{000} & $\alpha^5$\ts{0-10} $\alpha^3$\ts{00-1} & $\alpha^8$\ts{-1-10} $\alpha^{10}$\ts{-1-1-1}\\
        \hline
	$\alpha^1$ &$\alpha^2$\ts{000} $\alpha^4$\ts{000} & $\alpha^0$\ts{000} $\alpha^9$\ts{-100} &  $\alpha^3$\ts{000} $\alpha^{11}$\ts{-1-10} \\
        \hline
	$\alpha^2$ &$\alpha^{10}$\ts{-100} $\alpha^{11}$\ts{-100} & $\alpha^1$\ts{000} $\alpha^4$\ts{000} &   $\alpha^7$\ts{000} $\alpha^5$\ts{001} \\ 
        \hline
	$\alpha^3$ & $\alpha^0$\ts{001} $\alpha^5$\ts{0-11} & $\alpha^6$\ts{000} $\alpha^7$\ts{000} &  $\alpha^1$\ts{000} $\alpha^{11}$\ts{-1-10} \\
        \hline
	$\alpha^4$ & $\alpha^5$\ts{000} $\alpha^8$\ts{000}  & $\alpha^1$\ts{000} $\alpha^2$\ts{000}  & $\alpha^6$\ts{000} $\alpha^9$\ts{000} \\
        \hline
	$\alpha^5$ & $\alpha^2$\ts{00-1} $\alpha^7$\ts{00-1}& $\alpha^0$\ts{010} $\alpha^3$\ts{01-1} & $\alpha^4$\ts{000} $\alpha^8$\ts{000}  \\ 
        \hline
	$\alpha^6$ & $\alpha^3$\ts{000} $\alpha^7$\ts{000} & $\alpha^8$\ts{0-10} $\alpha^{11}$\ts{0-10} & $\alpha^4$\ts{000} $\alpha^9$\ts{000} \\ 
        \hline
	$\alpha^7$ & $\alpha^9$\ts{001} $\alpha^{10}$\ts{000} & $\alpha^3$\ts{000} $\alpha^6$\ts{000} & $\alpha^2$\ts{000} $\alpha^5$\ts{001} \\
        \hline
	$\alpha^8$ & $\alpha^6$\ts{010} $\alpha^{11}$\ts{000} & $\alpha^4$\ts{000} $\alpha^5$\ts{000} & $\alpha^0$\ts{110} $\alpha^{10}$\ts{00-1} \\ 
        \hline
	$\alpha^9$ & $\alpha^4$\ts{000} $\alpha^6$\ts{000} & $\alpha^0$\ts{100} $\alpha^1$\ts{100} & $\alpha^7$\ts{00-1} $\alpha^{10}$\ts{00-1} \\
        \hline
	$\alpha^{10}$ & $\alpha^0$\ts{111} $\alpha^8$\ts{001} & $\alpha^2$\ts{100} $\alpha^{11}$\ts{000} & $\alpha^7$\ts{000} $\alpha^9$\ts{001} \\
        \hline
	$\alpha^{11}$ & $\alpha^1$\ts{110} $\alpha^3$\ts{110}& $\alpha^8$\ts{000} $\alpha^6$\ts{010} & $\alpha^{10}$\ts{000} $\alpha^2$\ts{100}  \\ 
        \hline
\end{tabular}

\caption{\label{tab:hhk}Neighbours (nearest and next-nearest) of 12 sublattices ($\alpha^0 \ldots \alpha^{11}$) in the unit cell $(0,0,0)$ in the HHK network of corner-sharing triangles. The neighbours are grouped according to their participation in triangular plaquettes. Subscripts denote unit cell coordinates. }
\end{table}
\section{Matrix Cellular Automata}
\label{sec:CA}

In this section, we develop an understanding of the fractal structures in both the trillium and HHK  models in terms of matrix CAs. Both lattices are described by a cubic Bravais lattice with a basis. We  define a ``slice" to be spins of a particular sublattice at a fixed $z$-plane, \textit{i.e.,} with a certain value of the unit cell $z$-coordinate. The CA description follows from the possibility of grouping such slices into ``layers", such that in a ground state, the spin configuration of a layer uniquely determines the spin configuration of the two layer following and preceding it. 
\subsection{Trillium}

In trillium, we first group the spins into ``slices" according to their sublattice and $z$-coordinate as mentioned before, and order them as follows:

\begin{equation}
\ldots \beta_{-1}, \alpha_{-1}, \delta_{0}, \gamma_0, \beta_0, \alpha_0,\delta_1,\gamma_1 \ldots
\label{eq:levels}
\end{equation}
where the subscripts denote the $z$-coordinate of the unit cell. 
This ordering reveals the crucial structure that spins in each slice belong to triangular plaquettes whose  other spins lie entirely in the two slices preceding or following it. (Specifically, of the three plaquettes each spin participates in, one has its other spins entirely in the two previous slices, one has them  entirely in the next two slices, and the third plaquette has one of its other spins  in the previous slice, and one in the next slice.) As a result, if spins in any two consecutive slices are specified, the 
ground state constraints allow one to uniquely determine the spins of all other slices. 

On a cylinder (with $x$ and $y$ periodic), this implies the existence of $2^{2 L_x L_y}$ ground states. The ground states also determine the $\mathrm{log}_2 W$ fractal (subsystem) symmetry generators, where $W$ is the ground state degeneracy.

Given the spin configuration of any two consecutive slices, the spin configuration of the following slice can be determined using only the first plaquettes listed in Tab.~\ref{tab:trillium}:
\begin{align}
\nonumber s^\beta(i,j,k) &=s^\delta(i,j+1,k) s^\gamma(i-1,j,k)\\
\nonumber s^\alpha(i,j,k) &=s^\beta(i,j,k)s^\gamma(i,j,k)\\ 
\nonumber s^\delta(i,j,k+1) &=s^\alpha(i,j,k)s^\beta(i,j-1,k) \\ 
s^\gamma (i,j,k+1)&=s^\alpha(i,j,k)s^\delta(i+1,j,k+1). 
\label{eq:const_equations}
\end{align}    
Inverting these relations, it is easy to see that two consecutive slices of spins also determine the spin configuration of the slice preceding them. 
Without loss of generality, we describe ground states by specifying spin configurations in sublattices $\delta$ and $\gamma$ in a given layer.  

representing them by two polynomials over $\ff$,
\begin{align}
\nonumber u_k(x,y) &=\sum_{ij} s_{\delta}(i,j,k)x^i y^j \\
\nonumber v_k(x,y) &=\sum_{ij} s_{\gamma}(i,j,k)x^i y^j. 
\end{align}

We group two slices corresponding to sublattices $\delta$ and $\gamma$ into a layer, indexed by their $z$ coordinate.
If the spin configurations are specified on a layer $z=k$, the ground state spin configuration at $z=k+1$, obtained by Eq.~\eqref{eq:const_equations}, can be described by a matrix CA with a transition matrix $M(x,y)$

\begin{align}
\begin{pmatrix}
  u_{k+1}(x,y) \\
  v_{k+1}(x,y) 
\end{pmatrix}
&= M(x,y)
\begin{pmatrix}
  u_{k}(x,y) \\
  v_{k}(x,y) 
\end{pmatrix}, \\
M(x,y) &=
\begin{pmatrix}
1+\yb&1+x+xy\\
\xb+\xb \yb+\yb&x+\xb+y
\end{pmatrix},
\label{eq:CA}
\end{align}
where we have introduced the notation $\xb=x^{-1}$. Repeated application of $M$ allows us to determine $u_k(x,y)$ and $v_k(x,y)$ for all $k$. For convenience, we will gather $u_k(x,y)$ and $v_k(x,y)$ into a vector $\psi_k(x,y)= (u_k(x,y),v_k(x,y))^{T}$, so that
\begin{align}
  \psi_{k+p}(x,y)&=M^p(x,y) \psi_k(x,y) 
\end{align}

Note that in choosing to work with the $\delta, \gamma$ sublattices we have broken the symmetry of the lattice by choosing a preferred orientation for `slicing'. However, ground state configurations can be described in terms of any two consecutive slices in other orientations by closely related transition matrices, 
with  the same trace and determinant as those in  Eq.~\ref{eq:CA}.

Note that, in contrast to the transition function of the NM model, the transition matrix is \emph{invertible} and the equation above is valid for negative $p$. However, unlike polynomials over $\ff$, $M$ does not satisfy the `Freshman's dream', and hence the nature of fractonic excitations do not follow from the construction outlined for 
the NM model. For the same reason, the one-to-one mapping between spin and defect configurations which leads to an exact solution of the thermodynamics for the NM and square-pyramid models is also not available in the same way.

\subsection{HHK}
\label{sec:ca_hhk}

To construct the matrix CA describing the fractal structure on the HHK lattice, we first partition all sites of the lattice into layers $L_0(z), L_1(z), L_2(z)$ indexed by their $z$-coordinate. Each layer now consists of four slices (spins belonging to a particular sublattice at a certain value of $z$ coordinate)  :
\begin{align}
  \nonumber  L_0(z) &= \big\{ \mathbf{r} \in \{ \alpha^0_{xyz},\alpha^5_{xyz}, \alpha^8_{xyz}, \alpha^9_{xyz} \}, \forall xy \big\} \\
  \nonumber  L_1(z) &= \big\{\mathbf{r} \in \{ \alpha^1_{xyz},\alpha^4_{xyz}, \alpha^6_{xyz}, \alpha^{11}_{xyz}\}, \forall xy \big\} \\
  L_2(z) &= \big\{\mathbf{r} \in \{ \alpha^2_{xyz},\alpha^3_{xyz}, \alpha^7_{xyz}, \alpha^{10}_{xyz}\}, \forall xy \big\}.
  \label{eq:levels_hhk}
\end{align}
If we order the layers as 
\begin{align}
  \ldots L_0(z-1),L_1(z-1),L_2(z-1),L_0(z),L_1(z),L_2(z) \ldots,
\end{align}
then the ground state constraints uniquely determine the spin configurations of a layer from an adjacent layer.
Therefore, specifying the spin configuration in any single layer fixes the  ground state spin configuration of the entire system. For a system on a cylinder (with $x$ and $y$ directions periodic), this implies the existence of $W=2^{4L_x L_y}$ ground states, corresponding to the freedom of choosing spin configurations in any one layer. As before, the spin configurations of the $W$ ground states determine $\log_2 W$ generators of fractal symmetry.

Imposing ground state constraints on the plaquettes listed in Tab.~\ref{tab:hhk} uniquely determines the spins in each layer if the spins in the previous layer are specified. Spins in $L_0(k)$ determine those of $L_1(k)$ as follows:
\begin{align}
\nonumber  \ssl{1}(i,j,k)&=\ssl{0}(i,j,k)\ssl{9}(i-1,j,k),\\
\nonumber  \ssl{4}(i,j,k)&=\ssl{5}(i,j,k)\ssl{8}(i,j,k),\\
\nonumber  \ssl{6}(i,j,k)&=\ssl{4}(i,j,k)\ssl{9}(i,j,k),\\
  \ssl{11}(i,j,k)&=\ssl{8}(i,j,k)\ssl{6}(i,j+1,k).
  \label{eq:hhk_const1} 
\end{align}
$L_1(k)$ determines $L_2(k)$ as 
\begin{align}
\nonumber  \ssl{2}(i,j,k)&=\ssl{1}(i,j,k)\ssl{4}(i-1,j,k),\\
\nonumber  \ssl{10}(i,j,k)&=\ssl{2}(i+1,j,k)\ssl{11}(i,j,k),\\
\nonumber  \ssl{3}(i,j,k)&=\ssl{1}(i,j,k)\ssl{11}(i-1,j-1,k),\\
  \ssl{7}(i,j,k)&=\ssl{3}(i,j,k)\ssl{6}(i,j,k) .
  \label{eq:hhk_const2} 
\end{align}
Finally, $L_0(k+1)$ is determined by $L_2(k)$ as 
\begin{align}
\nonumber  \ssl{5}(i,j,k+1)&=\ssl{2}(i,j,k)\ssl{7}(i,j,k)\\
\nonumber  \ssl{9}(i,j,k+1)&=\ssl{7}(i,j,k)\ssl{10}(i,j,k)\\
\nonumber  \ssl{0}(i,j,k+1)&=\ssl{5}(i,j-1,k+1)\ssl{3}(i,j,k)\\
  \ssl{8}(i,j,k+1)&=\ssl{0}(i+1,j+1,k+1)\ssl{10}(i,j,k).
  \label{eq:hhk_const3} 
\end{align}

Inverting each of the relations in Eqs.~\eqref{eq:hhk_const1}--\eqref{eq:hhk_const3} tells us that the ground state constraints also determine spins in each layer from spins in the following layer. As usual, we describe spin configurations by the following polynomials over $\ff$:
\begin{align}
u^{q}_k(x,y)=\sum_{i,j}\ssl{q}(i,j,k)x^i y^j.
\end{align}

Without loss of generality, we can describe ground state configurations using the spins in layer $L_0(z), \forall z$, described by the polynomials $(u^0_k(x,y),u^5_k(x,y),u^8_k(x,u),u^9_k(x,y))$. Given the spin-configurations at $z=k$, the configurations at $z=k+1$ can be described by matrix CA, with a $4 \times 4$ transition matrix $M$ 

\begin{align}
&\begin{pmatrix}
  u^0_{k+1}(x,y) \\
  u^5_{k+1}(x,y) \\
  u^8_{k+1}(x,y) \\
  u^9_{k+1}(x,y) 
\end{pmatrix}
= M_0(x,y)
\begin{pmatrix}
  u^0_{k}(x,y) \\
  u^5_{k}(x,y) \\
  u^8_{k}(x,y) \\
  u^9_{k}(x,y) 
\end{pmatrix} 
\label{eq:CA_hhk0}
\end{align}
For convenience, we gather these polynomials into a vector $\phi^0_k(x,y)= (u^0_k(x,y),u^5_k(x,y),u^8_k(x,u),u^9_k(x,y))^T$. Similarly, if  we choose to describe ground states by the spins in layers $L_1(z)$ or $L_2(z)$, we have the vectors $\phi^1_k(x,y)= (u^1_k(x,y),u^4_k(x,y),u^6_k(x,u),u^{11}_k(x,y))^T$ and $\phi^2_k(x,y)= (u^2_k(x,y),u^3_k(x,y),u^7_k(x,u),u^{10}_k(x,y))^T$ respectively. Their CA evolutions would be similarly described by the matrices $M_1$ and $M_2$. In general, we have
\begin{align}
  \phi^i_{k+p}(x,y)=M^k_i(x,y) \phi^i_{p}(x,y).
  \label{eq:hhk_ca_evol}
\end{align}

Explicit forms for the matrices $M_0,M_1$ and $M_2$ can be determined from Eqs.~\eqref{eq:hhk_const1}--\eqref{eq:hhk_const3}. They are cumbersome and not directly relevant to the rest of the discussion, so we relegate them to Appendix.~\ref{app:hhk_matrices}. However, it will be crucial that the transition matrices satisfy the characteristic equation
\begin{align}
\nonumber  M^4_i&= T[M^3_i+M^1_i] +M^2_i[(T^2+(x+\xb)(y+\yb)] +1,\\
T&=\Tr[M_i]=x+\xb+y+\yb.
\label{eq:trillium_characteristic}
\end{align}

Finally, we note that the grouping of spins into layers described in Eq.~\eqref{eq:levels_hhk}, which enables the description of ground states by CA, is not unique. We show the consequences of  other choices in Appendix~\ref{app:hhk_partitions}.
\section{Fractonic excitations}
\label{sec:fractons}
For the NM model, the construction of fractonic excitations relies crucially on the CA transition function $f(x)$ following the Freshman's dream (Sec.~\ref{sec:review}). Specifically, if we begin with a configuration with a single defect and evolve with the CA transition matrix for $2^n$ steps, we arrive at a configuration with $3$ defects. This follows from the fact that $f(x)^{\tn}=1+x^{\tn}$, \textit{i.e.,} the spin configuration obtained by CA evolution for $\tn$ steps has only two flipped spins, independent of $n$. Although the matrix CAs which describe the trillium and HHK models (Sec.~\ref{sec:CA}) 
are known to generate self-similar structures~\cite{Gutschow_Nesme_Werner}, 
 this simplifying property is no longer available: iterating the transition matrix $M$  $\tn$ times generically does not lead to a configuration with only a few defects. To construct elementary fractonic excitations, we turn instead to a distinct simplification by leveraging the Cayley-Hamilton theorem: namely,   that the  matrix $M$ satisfies its own characteristic equation. This allows us to express $M^{\tn}$ in terms of lower powers of $M$, and use this fact to construct low energy excitations which are created by flipping by fractal subsets of spins of arbitrarily large size.
\subsection{Trillium}
\label{subsec:trillium_fractons}

\begin{figure*} 
	\includegraphics[width=0.4\columnwidth]{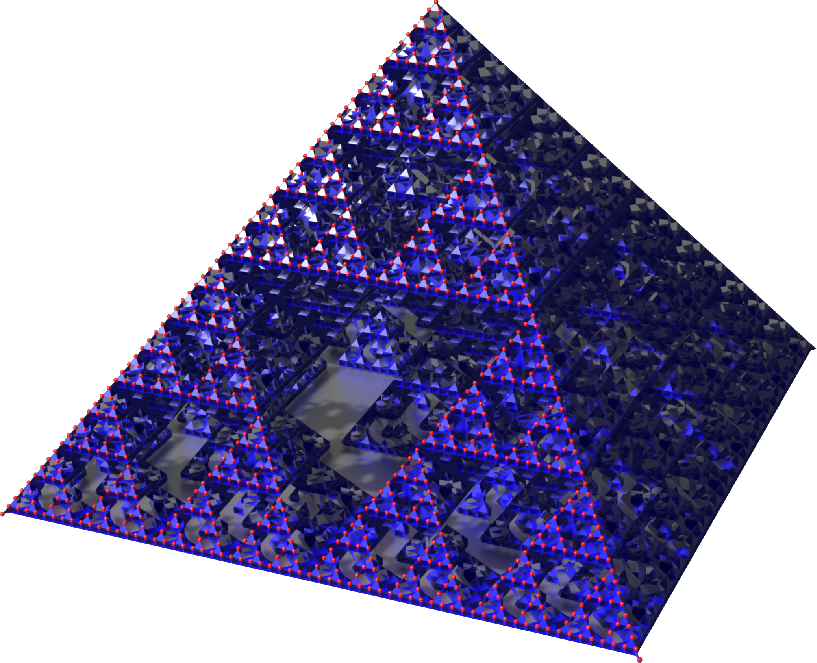}\hfill
	\includegraphics[width=0.4\columnwidth]{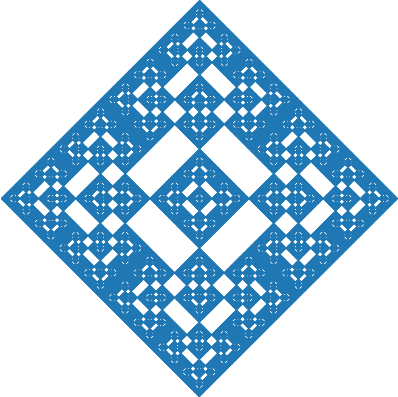}
	\caption{ The set of spins flipped to create an elementary fractonic cluster (EFC) for trillium is a fractal, with the fractons lying on the center and corners of a regular octahedron. Left: Flipped spins in the upper half of the octahedron. The flipped spins on the faces of the octahedron resemble the Sierpinski triangle (red dots). Right: Flipped spins in the plane  containing the center and 4 vertices of the octahedron. }
	\label{fig:trillium_fractal}
\end{figure*}

Following the strategy outlined above, 
we consider the characteristic polynomial $P(\lambda)$ of $M$,
\begin{equation}
P(\lambda)=\mathrm{det}M+\lambda\mathrm{Tr}M+\lambda^2.
\end{equation}
The Cayley-Hamilton theorem implies $P(M)=0$. Using  $\mathrm{det}M=1$ for the trillium CA (Eq.~\eqref{eq:CA}) we have  
\begin{alignat}{2}
      \nonumber     && M^2 &=M\cdot\Tr M + 1 \\
      \nonumber        &\implies& M &= \Tr M + M^{-1} \\
                       &\implies& M^{2^n} &= (\Tr M)^{2^n} + M^{-2^n}.
          \label{eq:trace_eq}
\end{alignat}
We multiply through by $M^{-1}$ to obtain the second line from the first, and square repeatedly using the properties of $\ff$ to obtain the third step from the second. From the above recursion relation we see that the trace plays an important role in the CA evolution, and consequently controls the fractal properties  of  the model. Explicitly,  the trace is given by an $\ff$ polynomial,
\begin{align}
\mathrm{Tr}M&=1+x+y+\bar{x}+\bar{y}\label{eq:explicittracetrill},
\end{align}
 and therefore satisfies a property similar to Freshman's dream, 
\begin{align}
(\Tr M)^{2^n}&=1+x^{2^n} + y^{2^n}+\xb^{2^n}+\yb^{2^n} \label{eq:tracepow},
\end{align}
From \eqref{eq:trace_eq} and  \eqref{eq:tracepow}, we see that $M^{2^n}$ is {\it almost} its own inverse, differing from it simply by the presence of additional spin flips at the locations specified by  \eqref{eq:tracepow}. This generalization of the Freshman's dream property to the matrix CA setting~\footnote{We resist the temptation to call this the  `Sophomore's Dream'.} motivates a route to constructing fractal excitations

We begin with a spin configuration at $\psi_{-2^n}(x,y)=(0,1)^T$, corresponding to a single down spin in the $\gamma$-sublattice at the unit cell $(0,0,-2^n)$. 
If all other spins with unit cell coordinate $z\leq -2^n$ are set to zero, then this initial configuration places a single defect on a $\alpha \delta \gamma$-plaquette. Now, applying the CA $2^n$ times in the $z$ direction yields the  $z=0$ spin configuration
\begin{equation}
\psi_{0}= M^{2^n}\psi_{-2^n} = (\Tr M)^{2^n}\psi_{-2^n} + M^{-2^n}\psi_{-2^n}.
\end{equation}
We consider the two terms on the RHS separately. As noted, from \eqref{eq:tracepow} we see that the first  term is similar to the `Freshman's dream' contribution in a scalar 
CA, and corresponds to 5 flipped $\gamma$-spins at $z=0$, with $xy$-coordinates $(0,0), (\pm2^n,0)$ and $(0,\pm2^n)$. 
We re-set each of these 5  spins to zero, thereby creating a defect in the $\alpha \delta \gamma$-plaquette below each spin. Using the properties of $\ff$ and Eq.~\eqref{eq:trace_eq}, we find that the resulting $z=0$ configuration is  
\begin{equation}
\bar{\psi}_0 = M^{2^n}\psi_{-2^n} + (\Tr M)^{2^n}\psi_{-2^n} = M^{-2^n}\psi_{-2^n},
\end{equation}
 corresponding to a macroscopic number of flipped spins. However, 
 we can perform a further evolution with $M^{2^n}$, so that the configuration at $z=2^n$, given by
\begin{align}
\nonumber  \psi_{2^n} &=M^{2^n}(\bar{\psi}_0)\\
\nonumber  &=M^{2^n}\big(M^{2^n}\psi_{-2^n} +(\Tr M )^{2^n} \psi_{-2^n} \big) &\\ 
\nonumber  &=M^{\tn}M^{-\tn}\psi_{-2^n}  &\text{(from Eq.~\eqref{eq:trace_eq})} \\
  &=\psi_{-2^n}=(0,1)^T
\end{align}
consists of a \textit{single} down spin, also on the $\gamma$-sublattice. Flipping this spin creates another defect in 
the $\alpha \delta \gamma$-plaquette below that spin. Thus we have constructed a fractonic excitation with 7 defects,  all of which lie on the $\alpha \delta \gamma$-plaquettes whose $\gamma$-spins are located at unit cell coordinates $(0,0,0),(\pm 2^n,0,0),(0,\pm 2^n,0),(0,0,\pm 2^n)$. The defects are thus placed at the corners and centres of a regular octahedron  whose corners are at distance $2^n$ from at the origin, where it is centred.  If one starts with initial conditions with a defect in either 
of $\alpha \beta \gamma$, $\beta \gamma \delta$, or $\delta\alpha \beta$-plaquettes (by starting with the vector $(0,1)^T$ in a description in terms of spins in consecutive layers $\beta,\alpha$; $\gamma,\beta$; or $\alpha,\delta$, respectively), following this construction creates 7 defects lying on plaquettes of the same type 
as the starting defect. This follows from the fact the transition matrices in these alternative descriptions have the same trace and determinant. For each of the four kinds of plaquettes, we have constructed  a series of fractonic excitations of energy $7J$, with the 7 defects 
living on the same kind of plaquettes at the corners  and center of  a regular octahedron whose corners lie $2^n$ unit cells away from its center.  We will call a set of excitations of this type an elementary fractonic cluster (EFC), and distinguish them from other multi-fracton configurations with higher energy.

Creating  EFCs involve flipping a fractal subset of spins. We can leverage the simplifications afforded by Eq.~\eqref{eq:trace_eq} to analytically determine the Hausdorff dimension $d_f$ of the generated fractals (Appendix ~\ref{app:fracdim}),
\begin{equation}
 d_f=\log_{2}((5+\sqrt{33})/2) \approx 2.43.
\end{equation}
The fractal subset of spins flipped to create an EFC  is depicted in Fig.~\ref{fig:trillium_fractal}. Following similar reasoning to the case of the NM model, it is evident that similar fractals (described by same the transition matrix $M$) can be used to describe flipped spins that correspond to distinct degenerate ground states.

\subsection{HHK}
\label{subsec:hhk_fractons}
We now use the CA presented in the previous section to construct EFCs of the Baxter-Wu model on the HHK network. As in our construction in Sec.~\ref{subsec:trillium_fractons}, we begin with the characteristic equation of $M_0$, given by
\begin{align}
  \nonumber  M_0^4 ={}&  T[M_0^3+M_0^1] \\ 
\nonumber        &  + M_0^2[(T^2+(x+\xb)(y+\yb)] +1 \\
\nonumber \implies  M_0^2+M_0^{-2}={}& T[M_0+M_0^{-1}] \\
                          &+[T^2+(x+\xb)(y+\yb)]\mathbb{I},  \label{eq:hhk_characteristic}\\ 
                              \text{where }T={}& \Tr(M)=x+\xb +y+\yb\nonumber.
\end{align}
Repeatedly squaring this equation gives us the recursion relation
\begin{align}	
    M_0^{2^{n+1}}+M_0^{-2^{n+1}}={} &  T^{\tn}[M_0^{2^{n}}+M_0^{-2^n}]\label{eq:hhk_characteristc2} \\
  & +[T^{2^{n+1}}+(x+\xb)^{\tn}(y+\yb)^{\tn}]\mathbb{I},\nonumber
\end{align}
which we solve to obtain
\begin{align}
  M_0^{\tn}+M_0^{-\tn}=
\begin{pmatrix}
  f^{\tn} & 0 & xy T^{\tn} & 0 \\
  \xb \yb T^{\tn} & f^{\tn} & T^{\tn} & T^{\tn} \\
  \xb \yb T^{\tn} & 0&  g^{\tn}  & 0 \\
   0 & T^{\tn}&  T^{\tn} & g^{\tn} \\
 \end{pmatrix} \label{eq:hhk_traceq},\\
\nonumber  \text{where } f=x+\xb, g=y+\yb, \text{when } n \text{ is odd},\\
\nonumber            f=y+\yb, g=x+\xb, \text{when } n \text{ is even}.
\end{align}
The characteristic equation.~\eqref{eq:hhk_characteristic} is also satisfied by the matrices $M_1$ and $M_2$ (Eq.~\eqref{eq:CA_hhk12}) which describe the CA evolution of spin-configurations of layers $L_1$ and $L_2$ respectively (Eq.~\eqref{eq:levels_hhk}). We can solve the corresponding recursions to obtain :
\begin{align}
\nonumber  M^{\tn}_1+M^{-\tn}_1=
\begin{pmatrix}
  f^{\tn} & 0 & 0& xy T^{\tn}  \\
  0 & g^{\tn}& T^{\tn} & 0 \\
  0 & T^{\tn}&  f^{\tn} & 0 \\
  \xb \yb T^{\tn} & 0&  0 & g^{\tn} \\
\end{pmatrix}\\
M^{-\tn}_2+M^{-\tn}_2=
\begin{pmatrix}
  g^{\tn} & 0 & T^{\tn}& 0  \\
  \yb T^{\tn} & g^{\tn}& 0 & xy T^{\tn} \\
  T^{\tn} & 0&  f^{\tn} & 0 \\
  \xb T^{\tn} & \xb \yb T^{\tn}&  \xb T^{\tn} & f^{\tn} \\
\end{pmatrix}\label{eq:hhk_traceq12} \\
\nonumber \text{where } f=x+\xb, g=y+\yb, \text{when } n \text{ is odd},\\
\nonumber            f=y+\yb, g=x+\xb, \text{when } n \text{ is even}.
\end{align}
As in the case of trillium, the fact that the matrix $M^{\tn}_i +M^{-\tn}_i$  depends on $n$ only through simple polynomials being raised to an exponent of $\tn$ immediately implies that the spin configurations generated through evolutions of  by $M^{\tn}_i$ and $M^{-\tn}_i$ differ by a few spins. This allows us to construct EFCs of arbitrarily large size with a fixed number of defects. 

  \begin{table}
	\caption{Table showing fracton configurations of elementary excitations  of size $\tn$. Elementary excitations have fractons on 8 different plaquettes lying on the corners of a regular octahedron whose vertices are $2^n$ lattice spacings away from its center. The notation $[(i),(j),(i,j)]$ denotes fractons on $A^i$ plaquettes at $(\pm \tn,0,0)$, $A^j$ plaquettes at $(0,\pm \tn,0)$ and both $A^i$ and $A^j$ plaquettes at $(0,0,\pm \tn)$.}
	\centering
	\begin{tabular}{|c||c|c|c|}
		\hline
		Level&  \multicolumn{3}{|c|}{Fracton configuration} \\
		\hline
		&Initial condition        & Even $n$ & Odd $n$ \\ 
		\hline \hline
		\multirow{4}{*}{$L_0$} & $(0,1,0,0)^T$ & $[(7), (4,7), (4)]$ &  $[(4,7), (7),(4)]$ \\
		&$(0,0,0,1)^T$ & $[(4,7), (4), (7)]$ &  $[(4), (4,7), (4)]$\\
		&$(0,1,0,1)^T$ & $[(4), (7), (4,7)]$ &  $[(7), (4,7), (4)]$\\
		&$(1,0,0,0)^T$ & $[(0,8), (8), (0)]$ &  $[(8), (0,8), (0)]$\\
		\hline
		
		\multirow{6}{*}{$L_1$} 
		&$(0,1,0,0)^T$ &  $[(6), ( 1,6), ( 1)]$      &    $[(1,6), ( 6), ( 1)]$\\
		&$(0,0,1,0)^T$ & $[(1,6), ( 1), ( 6)]$     &  $[(1), ( 1,6), ( 6)]$\\
		&$(0,1,1,0)^T$ & $[(1), ( 6), ( 1,6)]$     &  $[(6), ( 1), ( 1,6)]$\\
		&$(1,0,0,0)^T$ &  $[(0,8), ( 8), ( 0)]$ & $[(8), ( 0,8), ( 0)]$\\
		&$(0,0,0,1)^T$ &  $[(0), ( 0,8), ( 8)]$ & $[(0,8), ( 8), ( 0)]$\\
		&$(xy,0,0,1)^T$ & $[(8), ( 0), ( 0,8)]$ &  $[(0), ( 8), ( 0,8)]$\\
		\hline

		\multirow{4}{*}{$L_2$} 
		&$(0,1,0,0)^T$    & $[(2), ( 2,3), ( 3)]$&  $[(2,3), ( 2), ( 3)]$\\
		&$(0,0,0,1)^T$  & $[(2,3), ( 3), ( 2)]$  &  $[(3), ( 2,3), ( 2)]$\\
		&$(0,xy,0,1)^T$ & $[(3), ( 2), ( 2,3)]$  &  $[(2), ( 3), ( 2,3)]$\\
		&$(0,1,0)^T$    & $[(1,6), ( 1), ( 6)]$  &  $[(1), ( 1,6), ( 6)]$\\
		
		\hline
		
	\end{tabular}
	\label{tab:hhk_fractons}
\end{table}

The HHK model has many different kinds of EFCs. We explicitly show the construction of one kind of EFC, and only list the remaining cases for reasons of brevity. 

Consider a description in terms of the spins in $L_0$(Eq.~\eqref{eq:levels_hhk}) comprising the slices $(\alpha^0,\alpha^5,\alpha^8,\alpha^9)$. We start with an initial condition $\phi^0_{-\tn}=(0,1,0,0)^T$, which corresponds to a single flipped $\alpha^5$ spin in the unit cell $(0,0,0)$. We evolve the configuration by $M^{\tn}_0$ to get :
\begin{align}
  M^{\tn}_0 \phi^{0}_{-\tn}= M^{-\tn}_0 \phi^0_{-\tn} + (0,f^{\tn},0,T^{\tn})^{T}.
\end{align}

The second term corresponds to six down spins: two $\alpha^5$ spins and four $\alpha^9$ spins. We flip these spins back, and the remaining configuration is described by the first term $M^{-\tn}_0 \phi^0_{-\tn}$. Under further evolution by $M^{\tn}_0$, this evolves back to $\phi^0_{\tn}=\phi^0_{-\tn}=(0,1,0,0)^T$--- which describes one flipped $\alpha^5$ spin at $z={\tn}$. The resulting configuration has no defects except at the $\alpha^5$ spins at $(0,0,\pm \tn)$, the two $\alpha^5$ spins described by $f^{\tn}$ and the four $\alpha^9$-spins described by $T^{\tn}$ at $z=0$. Each of these eight spins have either one or two defect plaquettes adjacent to them. Flipping the ones which have two defect-plaquettes gives us an 8-defect excitation. This EFC has 8 defects, or fractons, lying at the corners (but not the center) of a regular octahedron, and is created by flipping a fractal subset of spins within that octahedron. For odd $n$, this quasiparticle has fractons on $A^{7}$-plaquettes at $(\pm \tn,0,0)$, on both $A^{7}$ and $A^4$-plaquettes at $(0,\pm \tn,0)$ and on $A^{4}$-plaquettes at $(0,0,\pm \tn)$. We schematically denote the location and nature of fractons of a such an EFC by the shorthand $[(7),(4,7),(4)]$. By beginning with different initial conditions (different  flipped spins on the $z=-2^n$ layer) we may construct different EFCs with different fractonic configurations. While constructing all the different EFCs  requires us to consider all  three matrices $M_i$ corresponding to different layers $L_i$ and  initial vectors $\phi^i_{0}$, in each case the procedure for their construction is the same:
\begin{enumerate}
  \item Choose an initial condition $\phi^i_{-\tn}$ at $z=-\tn$
  \item  Evolve with $M^{\tn}_i$ to obtain $\phi^i_{0}=M_i^{-\tn} \phi^i_{-\tn} +(M^{\tn}_i+M^{-\tn}_i)\phi^i_{-\tn}$.
  \item We infer from the form of $M^{\tn}_i+M^{-\tn}_i$ (Eqs. \eqref{eq:hhk_traceq} and \eqref{eq:hhk_traceq12})  that the second term describes a set of down spins such that the size of the set does not depend on $n$. We flip these spins back.
  \item We evolve the resulting spin configuration by $M^{\tn}_i$ to obtain $\phi^i_{\tn}=\phi^{i}_{-\tn}$.
   \item The resulting configuration has  an $n$-independent number of defects at $z=0$ and $z=\pm \tn$. If two defect-carrying plaquettes share a site, then we flip the spin on that site.
  \end{enumerate}

  We tabulate the initial conditions and CA matrices which create EFCs under this procedure  in Tab.~\ref{tab:hhk_fractons}, along with the locations of the  fracton configurations of such EFCs. 
  We note that each EFC involves $8$ fractons spread across two different kinds of plaquettes. In such EFCs, fractons are never hosted on plaquettes $A^5,A^9, A^{10}$ and $A^{11}$. These defects correspond to triads of next-nearest neighbours, in contrast to the other plaquettes which correspond to triads of nearest neighbours. The defect configurations of such EFCs of a given size $2^n$ have a sense of ``chirality"---e.g.~for odd $n$, there exist EFCs of size $2^n$ which have the fracton configurations $[(4), (4,7),(7)]$, $[(4,7), (7),(4)]$ and $[(7),(4),(4,7)]$, but no such EFC with fracton configuration $[(4),(7),(4,7)]$. This is a consequence of the dependence of Eqs.~\eqref{eq:hhk_traceq} and ~\eqref{eq:hhk_traceq12} on the parity of $n$.

  The fractal dimension of the set of spins flipped to create these EFCs can be calculated exactly, using simplifications following from  Eq.~\eqref{eq:hhk_characteristc2}.   The calculation closely follows the one used for trillium, and is explained in Appendix.~\ref{app:fracdim}; we find that   \begin{equation}
    d_f=\log_2(6) \approx 2.58.
  \end{equation}

\section{Trivial Thermodynamics and Glassy Dynamics} 
In this section, we use a combination of matrix CA technology developed over the preceding sections and numerical simulations to argue that both trillium and HHK display two key features in common with the 2D NM model: namely, their thermodynamics is trivial and can be mapped to that of the free defect gas, whereas their dynamics is glassy with logarithmic energy barriers. However, the arguments that allow a computation of the energy barriers is significantly more complex than for NM, as we explain below.

\label{sec:thermodynamics_and_glassiness}
\subsection{Thermodynamics}
\label{subsec:thermodynamics}
\begin{figure} 
    \includegraphics[width=0.8\columnwidth]{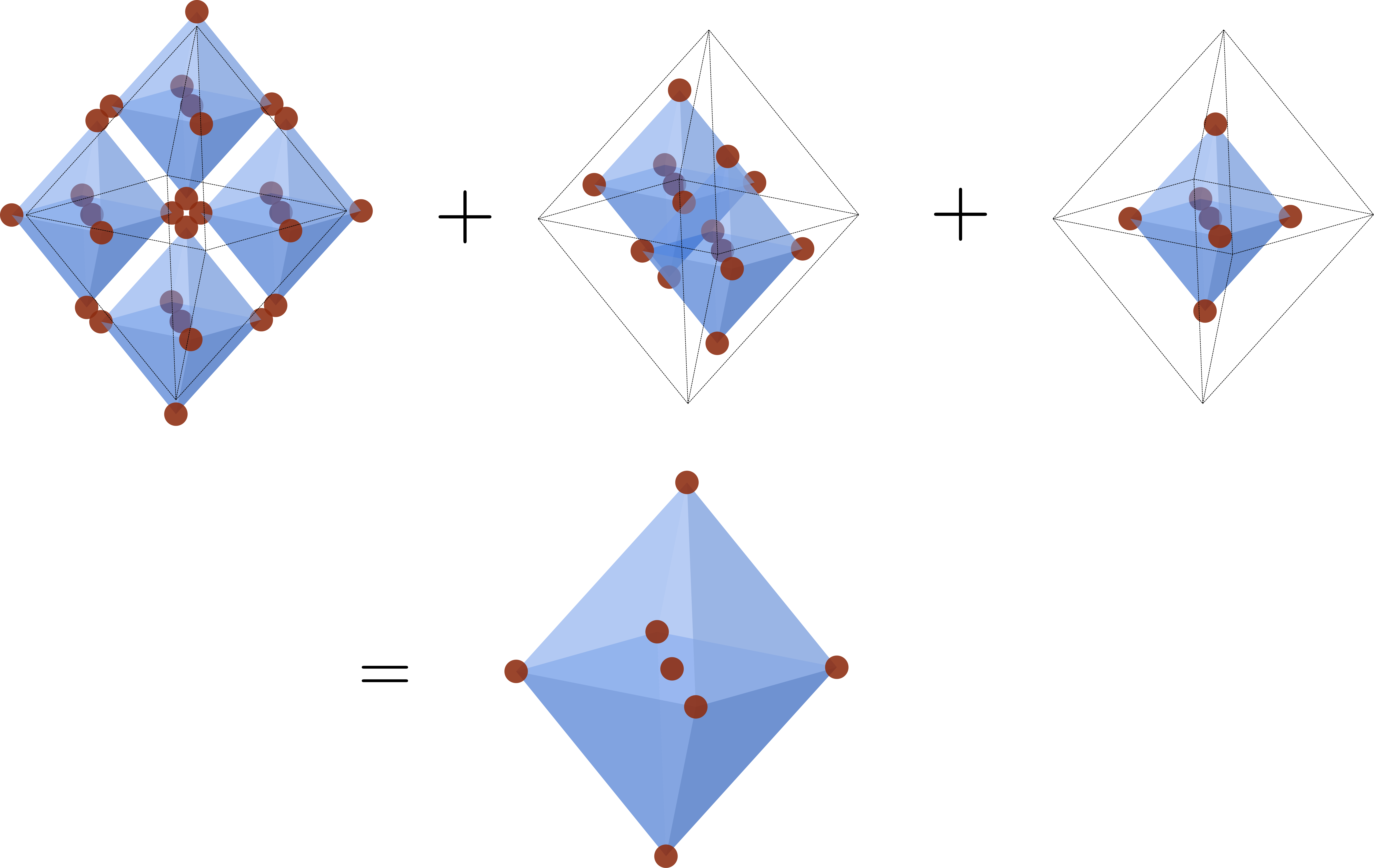}
    \caption{For trillium, 7 different EFCs of size $2^n$ fuse to form a EFC of size $2^{n+1}$, i.e., an EFC of size $2^{n+1}$ can be created by flipping the spins corresponding to the creation of each of the 7 smaller EFCs in series. For each of this excitations, all 7 fractons  are hosted by plaquettes of the same type. Fractons are denoted by solid red dots.}
\label{fig:trillium_fusion}
\end{figure}
\begin{figure} 
    \includegraphics[width=0.8\columnwidth]{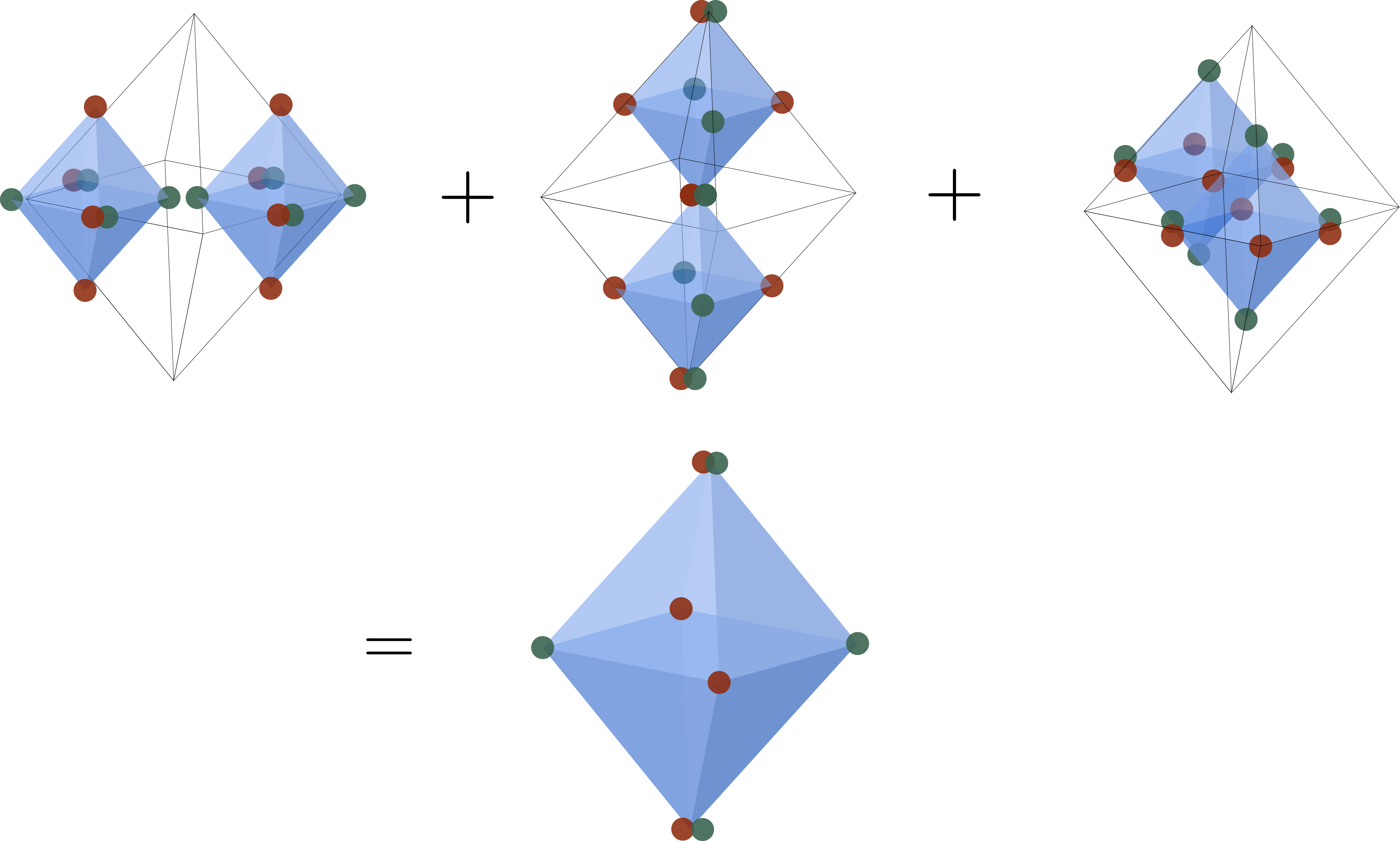}
    \caption{For HHK, 6 EFCs of size $2^n$ fuse to form an EFC of size $2^{n+1}$. A larger excitation can be created by flipping the spins corresponding to the creation of each of the 6 smaller excitations in series. In each of these excitations, fractons are hosted on 2 different kinds of plaquettes (\ref{tab:hhk_fractons}). The 2 kinds of fractons are denoted by the red and green dots.  }
\label{fig:hhk_fusion}
\end{figure}

In the case of trillium, we can construct a one-to-one mapping between spin and defect configurations for certain system sizes, as in the NM model. This  allows us to  exactly compute the partition function and hence solve the thermodynamics for these system sizes. In the NM model, the Freshman's dream allows us to prove that on tori of size $2^n$, the ground state condition is only satisfied by the unique spin configuration with all spins pointing up. Consequently on these system sizes there is a one-to-one mapping to the free defect problem, whence the partition function follows. The generalization to situations where the fractal structure is described by a matrix CA is not obvious. However, for trillium, the construction of 7-fracton excitations  in Sec.~\ref{subsec:trillium_fractons} provides us with a route past this problem. Consider a system on a torus with $2^{n+1}$ sites in each direction labelled by $-2^n,\ldots,2^n-1$, where $x^{2^n}$ is identified with
$x^{-2^n}$ (and similarly for $y$ and $z$); in this case, $(\Tr M)^{2^n}=1$ and consequently, Eq.~\ref{eq:trace_eq} simplifies to $M^{2^n}=M^{-2^n}+1$. Repeating the construction of the  EFC  as before creates a 
single $\beta\gamma\delta$-defect  at $(0,0,0)$, since the other six defects annihilate in pairs due to the periodic boundary conditions.

Similarly, starting from different initial conditions one can create configurations which host a single defect in any given plaquette of the system. 
By superposing spin-configurations (mod $2$) of different defects one can construct all possible defect configurations. Since the number of plaquettes (which can host defects) 
is equal to the total number of spins, the existence of all possible defect configurations implies that there is a one-to-one mapping between spin  and defect configurations for tori of size $2^{n+1},n\in \zz_{+}$. 
Since the defect variables are non-interacting, the partition function for a system with $N$ plaquettes is then given by
\begin{align}
    Z=[2\cosh(1/T)]^{N}.
\label{eq:free_defect}
\end{align}
leading to an average energy per plaquette  $\langle E\rangle=1/(1+\exp(1/T))$. The  absence of any thermodynamic phase transition is evident from the triviality of the partition function. While this result is exact for systems on tori of sizes which are powers of $2$, the corrections for other 
system sizes are sub-extensive and therefore are negligible in the thermodynamic limit. The argument (originally given for the NM model but  repeated here for completeness) is as follows~\cite{Newman_Moore}. For systems on tori which are not of size $2^{n+1}$, not all defect configurations are allowed; each allowed defect configuration corresponds to a multiplicity $\Omega$ of spin configurations, where $\Omega$ is the number of fractal symmetry generators, or equivalently, the number of ground states. However, all triangular plaquettes can independently host a defect, except possibly the last layer of plaquettes whose defect states are fixed by periodic boundary conditions. The partition sum can be now expressed in terms of defect states of the remaining  $N'$ plaquettes as $Z= \Omega \sum^{N'}_{n} {N'\choose  n}\exp(-(n+\delta n))$, where $\delta n$ is the number of defect plaquettes in the last plane which are fixed by periodic boundary conditions. However,  $\delta n<L_x L_y$ and consequently its contribution to the free energy and its derivatives are sub-extensive. For systems on a cylinder, the partition sum is simpler, $Z= \Omega \sum^{N}_n {N\choose  n} \exp(n)$. In both cases, the thermodynamic free energy per spin is identical to that obtained  from the free defect partition function of Eq.~\eqref{eq:free_defect}.

For the HHK model, there is no such one-to-one mapping between spin and defect configurations. We can show this by explicitly constructing 16 unit cell periodic spin configurations, all of which correspond to ground states. Therefore, irrespective of boundary conditions, there are at least 16 ground states, ruling out the possibility of an one-to-one mapping between spin and defect configurations. Such periodic ground states are tabulated in Appendix~\ref{app:hhk_periodic_ground state}. Despite the lack of a one-to-one mapping between spin and defect configurations, the arguments outlined in the previous paragraph lead us to expect bulk thermodynamic quantities to be well-described by the free defect partition function of Eq.~\eqref{eq:free_defect}. This can be verified by comparing the  energy computed from such a partition function against that  obtained from an unbiased Monte Carlo calculation (using a cluster algorithm to overcome slow equilibration times implied by the glassiness of the single-spin-flip dynamics). Such a comparison is shown in Fig.~\ref{fig:hhk_annealing}.

\subsection{Energy barriers and glassiness}
\begin{figure} 
    \includegraphics[width=0.8\columnwidth]{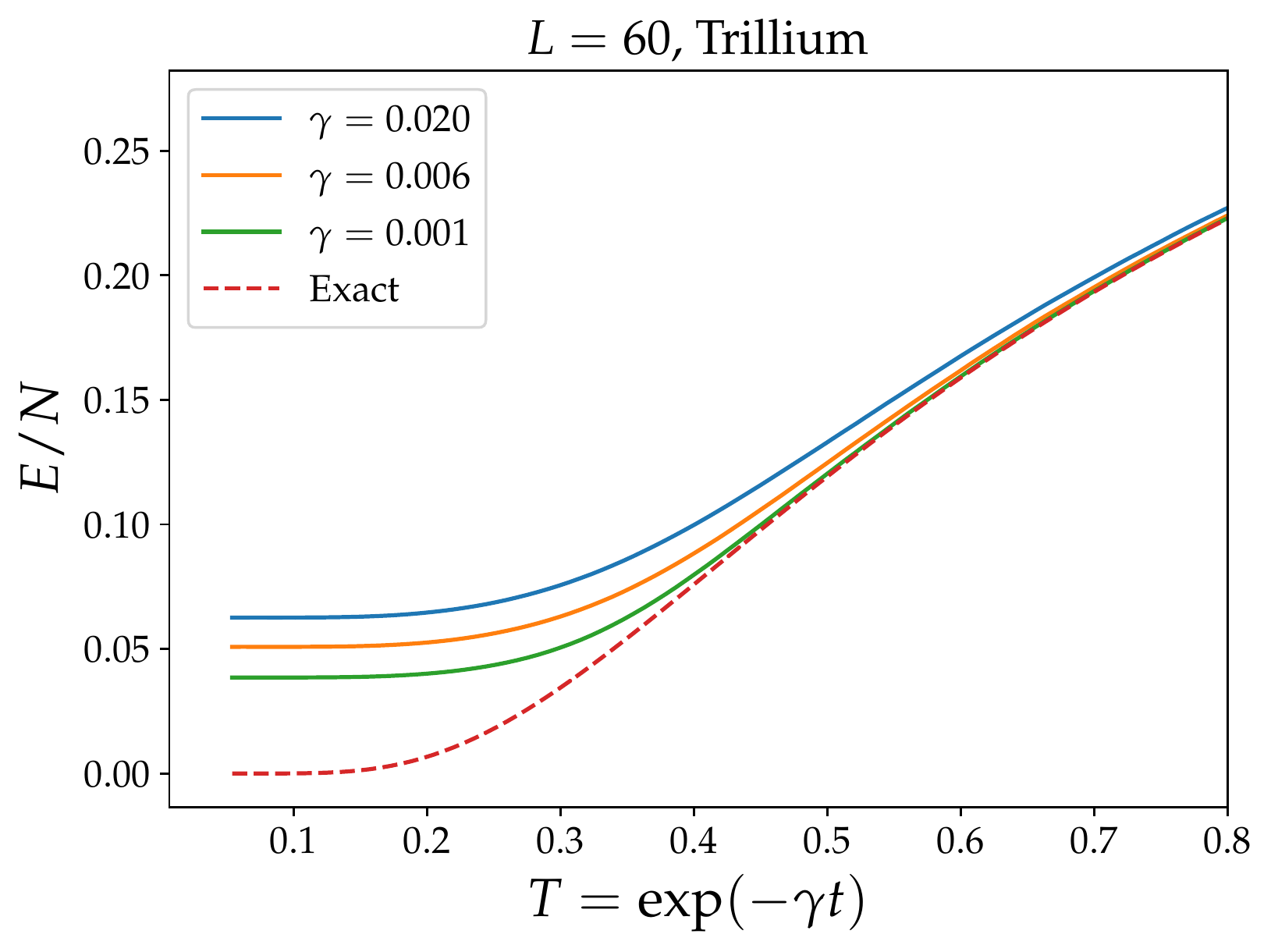}
    \caption{Relaxation of energy for annealing simulations in trillium. The temperature is cooled with as $T=\exp(-\gamma t)$. The measured value of energy stops tracking the analytical value of equilibrium energy, signalling a failure to reach equilibrium, at a temperature  $T_g(\gamma)$. $T_g$ increases with $\gamma$---a signature of glassiness.  }
\label{fig:trillium_annealing}
\end{figure}
\begin{figure} 
    \includegraphics[width=0.8\columnwidth]{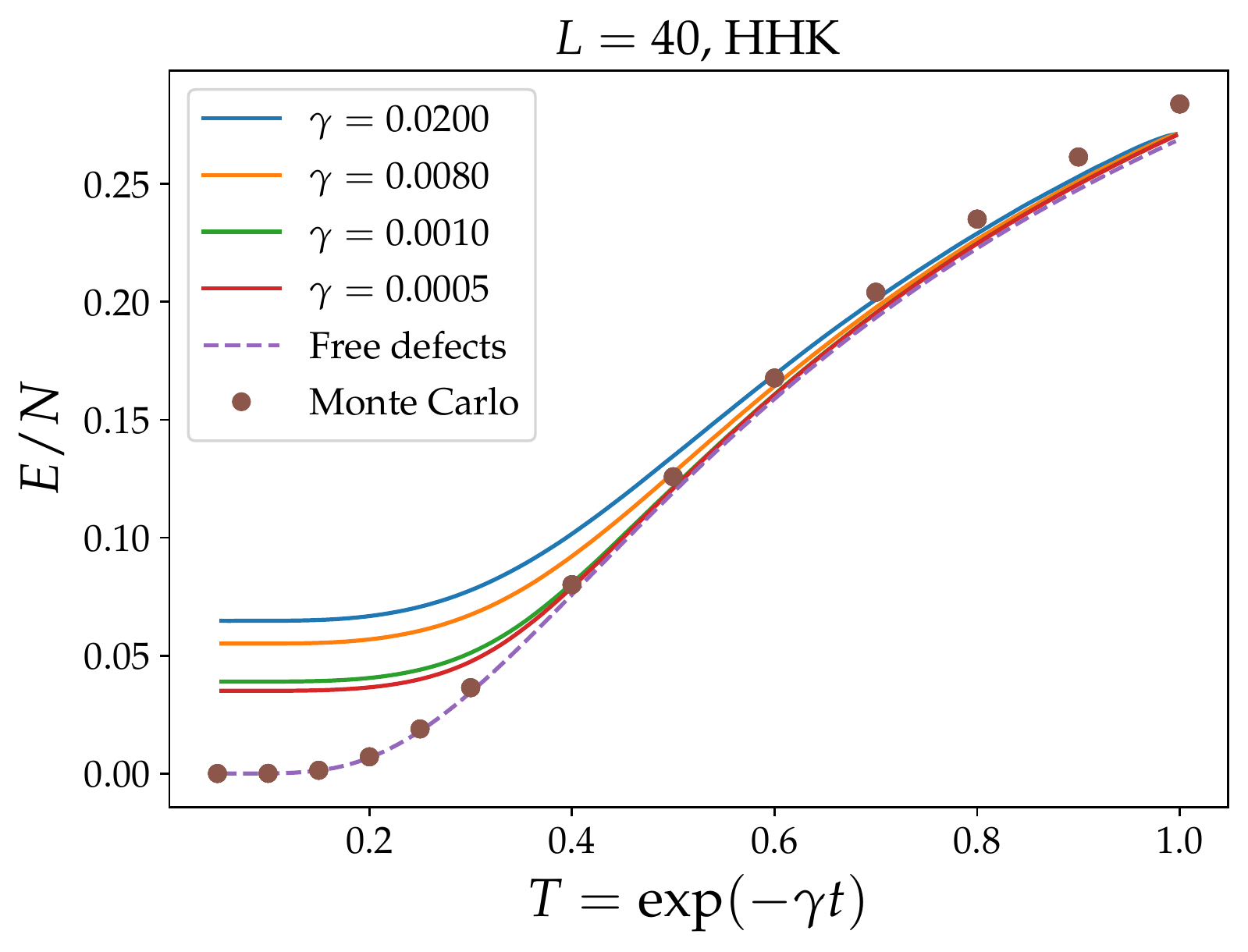}
    \caption{Relaxation of energy for annealing simulations in HHK. The temperature is cooled with as $T=\exp(-\gamma t)$. Like trillium, the measured value of energy stops tracking the equilibrium energy at a cooling-rate dependent temperature  $T_g(\gamma)$, which increases with $\gamma$. While an exact solution of equilibrium energy is not available, we see that the energy obtained by calculation for a gas of free defects (Eq.~\eqref{eq:free_defect}) agrees well with the energy obtained form a cluster Monte Carlo calculation.}
\label{fig:hhk_annealing}
\end{figure}
\label{subsec:glassiness}
To explore the possibility of glassiness in classical dynamics, it is important to determine the activation energies (`barriers') associated with creating or annealing away the EFCs. The origin of glassiness  often lies in the divergent scaling of these activation energies  with the system size. 
We will now argue that for both trillium and HHK, the relevant  barriers scale logarithmically with the linear system size. Essentially in both cases this follows from the observation that the spin configuration of an EFC of size $2^{n+1}$ is a superposition of the spin configurations $(\textrm{mod }2)$  of a fixed number (independent of $n$) of smaller EFCs of size $\tn$.  

In the case of trillium, as explained in Sec.~\ref{subsec:trillium_fractons}, each EFC has 7 fractons at the corners and center of a regular octahedron of size $2^n$,whose corners lie $2^n$ lattice spacings away from its center. Such EFCs involve spin-configurations with a fractal set of down spins. The important observation  here is that the configuration of such an EFC  of size $2^{n+1}$ centered at a unit cell $O$ is equal to the superpositions of spin-configurations $(\text{mod} 2)$ of 7 EFCs of size $2^n$--- with the smaller EFCs centred at the corners and center of an octahedron, centered at $O$,  of size $2^n$.  This can be verified explicitly, by constructing  the defect configuration resulting from the superposition and seeing that it indeed corresponds to the larger fracton. A schematic of a larger EFC being built out of superpositions of smaller EFCs is displayed in Fig.~\ref{fig:trillium_fusion}. 

For the HHK, each EFC has 8 fractons spread across two different types of plaquettes, all of them lying at corners (but not the center) of a regular octahedron of size $2^n$. Two corners of the octahedron hosts two defects, while the other six host one defect each. Without loss of generality, let us consider an EFC which host defects on two kinds of plaquettes of type $A^{i}$ and $A^{j}$, ones which have defect configurations $[(i),(j),(i,j)]$, $[(i,j),(i),(j)]$ and $[(j),(i,j),(i)]$. In the shorthand introduced in Sec.~\ref{subsec:hhk_fractons}, the configuration $[(i),(j),(i,j)]$ denotes defects on $A^i$ at positions $(\pm \tn,0,0)$, $A^j$ at positions $(0,\pm \tn,0)$ and on both $A^i$ and $A^j$ at positions $(0,0,\pm \tn)$. Now, an EFC of size $2^{n+1}$ with fracton configuration $[(j),(i),(i,j)]$ is a superposition $(\text{mod } 2)$ of six EFCs of size $\tn$: {(i)} two EFCs with fracton configuration $[(j),(i,j),(i)]$ centered at $(\pm \tn,0,0)$, {(ii)} two EFCs with fracton configuration $[(i,j),(i), (j)]$ centered at $(0, \pm \tn,0)$ and {(iii)} two EFCs of fracton configuration $[(i),(j),(i,j)]$ centered at $(0,0,\pm \tn)$. Superposing these EFCs in different ways also creates EFCs of size $2^{n+1}$ with fracton configurations  $[(i),(i,j),(j)]$ and $[(i,j), (j), (i)]$ . A schematic of such a larger EFCs being built out of superpositions of 6 smaller EFCs is shown  in Fig.~\ref{fig:hhk_fusion}. Note that the smaller EFCs have a chirality opposite to that of the larger EFC, consistent with the comments about chirality at the end of Sec.~\ref{subsec:hhk_fractons}.

For both HHK and trillium, creating an EFC involves flipping a fractal subset of spins from the ground state. The above results imply that an EFC in trillium (HHK)  of size $2^{n+1}$ can be created by sequentially flipping the spin configurations corresponding  to the 7 (6)  smaller EFCz of size $\tn$.
Although this guarantees that the barriers associated with an EFC of linear size $\ell \sim 2^n$ can grow at most as $n$ (i.e. logarithmically), the actual computation of barriers is more involved than for the NM model, as we now describe.

We estimate barriers associated with EFCs of size $2^{n+1}$ from barriers associated with excitations of size $\tn$.  For specificity, let us focus on the activation barriers associated with  EFCs in trillium. We start with EFCs of size $1$. For trillium, this involves $11$ flipped spins (for the HHK this involves $22$ flipped spins). The number of spin flips for a quasiparticle of size $1$ are small enough for an exhaustive search of all possible sequences of spin flips for the pathway which involves intermediate states of the lowest energy. This gives us the activation energy for creating, or annealing away, an EFC of size $1$, which consists of 7 fractons on 7 plaquettes at the corners and center of an octahedron. We will also need the activation energy associated with performing these spin flips when the initial conditions correspond to a different combination of these 7 plaquettes hosting fractons, since the associated   processes will correspond to intermediate steps when creating larger EFCs. We can now estimate the activation energy associated with creating an EFC of size $2$ by sequentially flipping the spins corresponding to the 7 smaller elementary excitations of size $1$. To look for the pathway corresponding to lowest energy, we search the $7!$ different orderings of flipping the spins corresponding to the 7 smaller EFCs. It is clear that this would require the activation energy of flipping the spins corresponding to the fracton of size $1$ with various combinations of defect-states on its corners and center, calculated in the previous step. We can iterate this procedure and use the activation energy associated with excitations of size $2^n$ to calculate the activation energy of an EFC of size $2^{n+1}$. Doing so, we find that activation energy $E_A(n)$ of an EFC of size $2^n$ is given by 
\begin{align}
  \nonumber E_A(n)&=f_n+E_A(n-1) \\
  \nonumber E_A(0)&=J\\
  f_n&= \begin{cases}
    8J, &\text{for }  $n=2,4,5,7,8,10,11 \ldots$ \\
    7J, &\text{for }  $n=1,3,6,9,12 \ldots$ \\
  \end{cases}
  \label{eq:trillium_barriers}
\end{align}
Note that  the NM model enjoys the  feature that that energy barrier of a defect of size $2^{n+1}$ only differs from that of a defect of size $2^n$ due to the extra energy of a single additional plaquette. This obviates the need for the exhaustive search over combinatorial pathways and thereby simplifies the barrier computation.

\begin{figure}
    \includegraphics[width=\textwidth]{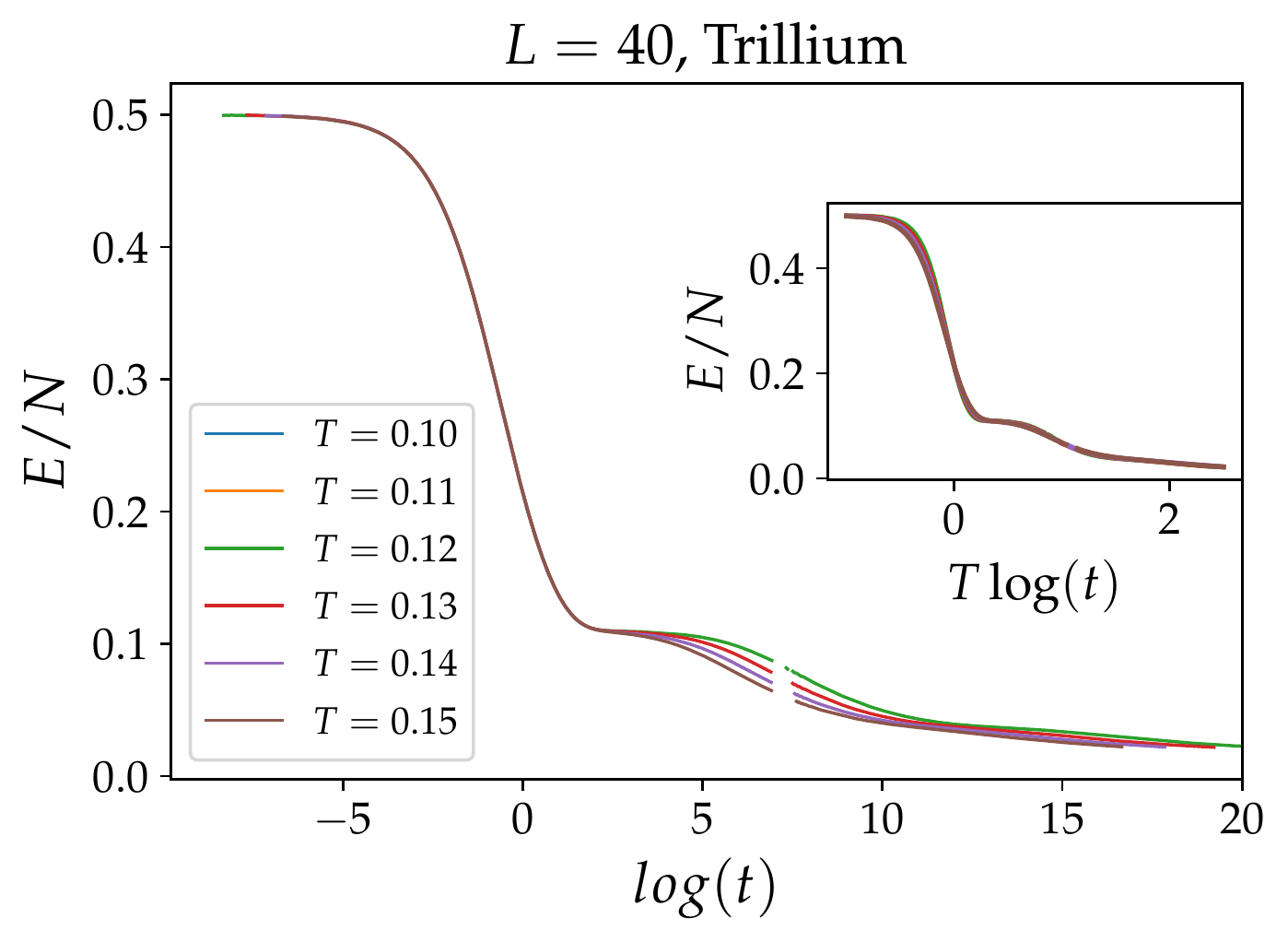}
    \caption{\label{fig:trillium_dynamics}Time evolution of intensive energy  $E/N$, for trillium. Like trillium, the plateaus correspond to different stages of the hierarchical dynamics. The different plateaus visible in the figure correspond to stages of the dynamics where higher energy excitations are being annealed away. The curves approximately collapse when plotted against $T \log(t)$, as the mean defect spacing $d_m \sim t^T$ is the dominant lengthscale in the problem. }   
  \end{figure}
\begin{figure}
    \includegraphics[width=\textwidth]{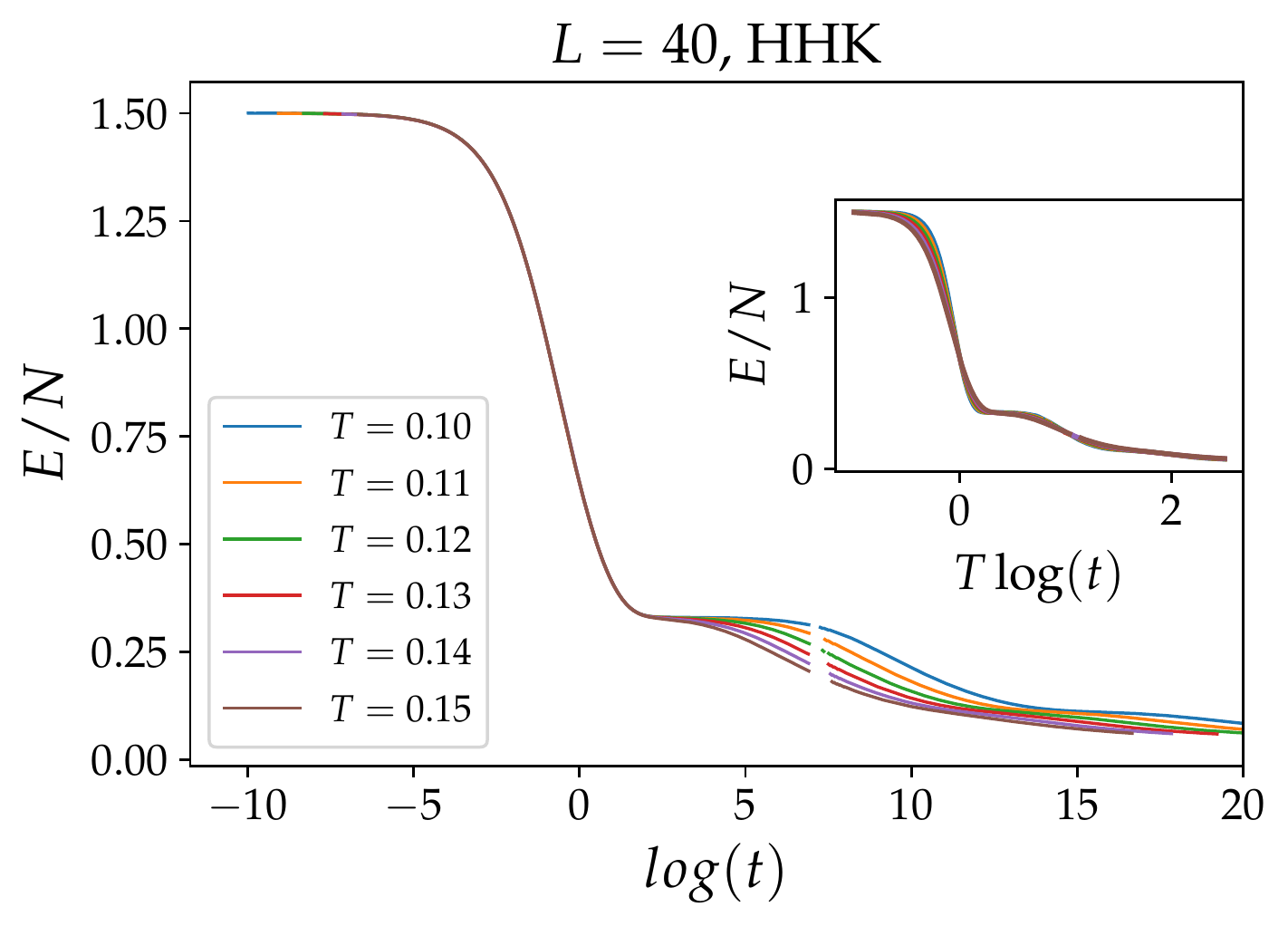}
    \caption{\label{fig:hhk_dynamics}Time evolution of intensive energy  $E/N$, for HHK. The plateaus correspond to different stages of the hierarchical dynamics. Note that the different stages do not correspond to stages where EFCs of different sizes are annealed away, but are stages where excitations of higher energy (also fractonic) are annealed away. The curves approximately collapse when plotted against $T \log(t)$, as the mean defect spacing $d_m \sim t^T$ is the dominant lengthscale in the problem. }   
  \end{figure}

 We repeat a similar calculation for HHK and obtain barriers given by
\begin{align}
  \nonumber E_A(n)&=f_n+E_A(n-1) \\
  \nonumber E_A(0)&=J\\
  f_n&= \begin{cases}
    7J, &\text{for }  n=2,3. \\
    6J, &\text{for }  n \notin \{2,3\}  \\
  \end{cases}
  \label{eq:hhk_barriers}
\end{align}
Thus, for both systems, the barriers associated with EFCs of size $2^n$ grow as $n$.
    
A system of linear dimension $L$ can host EFCs of a similar size; using $L\sim 2^n$  and the above arguments we conclude that such a system must overcome barriers of $k_L J\log _2(L)$ to equilibrate, where $k$ has a very weak and non-monotonic $L$ dependence for trillium (Eq.~\eqref{eq:trillium_barriers}). This logarithmic scaling implies  super-Arrhenius behaviour of relaxation times, which follows from similar arguments as to those used in the NM model~\cite{Garrahan_Newman}, which can be summarized as follows. The timescale for relaxing a EFC of size $d$ is given by the Arrhenius formula $t_d \sim \exp( kJ\log_2 (d)/T) $. After the system has relaxed 
for a time $t$, we expect the mean defect spacing $d_m \sim t^{T/kJ\log(2)}$. The equilibrium solution of the energy tells us $d_m \sim \langle E/N \rangle ^{-1/3} \sim \exp(1/3T)$, 
and consequently an equilibration time of
\begin{equation}
  t_{\mathrm{eq}} \sim \exp \Big(\frac{1}{3 T^2 \log(2)}\Big).
  \label{eq:eits}
\end{equation}
Such super-Arrhenius behaviour of the relaxation timescale characterises what has been termed \textit{fragile glassiness}, to be contrasted with  strong, Arrhenius, behaviour where $ t_{\mathrm{eq}}\sim \exp(1/T)$~\cite{Debenedetti_Nature,Angell_Science}. Fragile glassiness points to the presence of  barriers to relaxation which grow with decreasing temperature.

We numerically explore the single-spin dynamics and its glassy behaviour using the Bortz-Kalos-Lebowitz algorithm~\cite{Bortz_Kalos_Lebowitz} on 3D samples of size $4\times L \times L \times L$. First, we perform annealing simulations where we cool the system from high temperatures as 
$T=\exp(-\gamma t)$ with different cooling rates $\gamma$. We display the resulting time dependence of energies for both systems in Fig.~\ref{fig:trillium_annealing} and Fig.~\ref{fig:hhk_annealing}. The temperature $T_g$ where the system falls out of equilibrium  (where the energy fails to track to the exact expression) is dependent on the cooling rate---a classic signature of glassiness~\cite{Cugliandolo, Debenedetti_Nature}. 
If we plot the intensive energy $E/N$ against the time elapsed (Figs.~\ref{fig:trillium_dynamics} and ~\ref{fig:hhk_dynamics})  at different temperatures, we see plateaus which reveal metastable states characterising the hierarchical nature of the dynamics. 

Since the mean defect separation  $d_m \sim t^{T/k\log(2)}$ is the dominant lengthscale, 
the curves approximately collapse when plotted against $d_m\sim t^T$ (as shown in the inset).  However, unlike the NM model, each plateau does not characterize a stage of the dynamics where EFCs of a given size are annealed away. Instead, they represent stages where fractonic excitations of higher energy are annealed away. These excitations are associated with intermediate stages of annealing away an EFC.

\section{Quantum Fluctuations}

\subsection{Introducing a transverse field} 
\label{subsec:multispin_corrs}
While the solution of the thermodynamics (Eq.~\eqref{eq:free_defect}) shows the absence of classical phase transitions, we can  instead  drive a quantum phase transition by introducing quantum fluctuations via a transverse field term, 
\begin{equation}
H_{\mathrm{TF}}=-J\sum_{\Delta}\sigma^z_i \sigma^z_j \sigma^z_k-h\sum{\sigma^x_i}.
\end{equation}
$H_{\mathrm TF}$ is amenable to duality transformations: we map the energy term 
on each triangular plaquette $a$ to a Pauli spin $\tau^x_a \rightarrow \sigma^z_i \sigma^z_j \sigma^z_k$, and the transverse field term (which flips three spins) is transformed as 
$\sigma^x_i \rightarrow \tau^z_a \tau^z_b \tau^z_c$, where $a$, $b$ and $c$ are the plaquettes which have the site $i$ as their shared corner. 
This transformations map ${H}_{\mathrm{TF}}$ to $\tilde{H}_{TF}$, given by 
\begin{equation}
\tilde{H}_{TF}=-h\sum_{i}\tau^z_a \tau^z_b \tau^z_c-J\sum_a{\tau^x_a}.
\end{equation}
Since the trillium lattice is self-dual, 
it follows that  ${H}_{\mathrm{TF}}$ is self-dual. Therefore, if it exhibits a single phase transition it must occur at $J=h$. We expect such a transition to be a candidate for a novel critical point with spontaneous fractal-symmetry-breaking, as was recently observed in the simpler case of the transverse-field NM model~\cite{YizhiEA} in 2D. While the corresponding model on HHK is not self-dual, we still anticipate a similar phase transition at a $h_c \sim J$.
We can employ standard methods~\cite{TrithepEA,YizhiEA} to construct many-body correlation functions that diagnose the ordered phase and phase transitions.  The relevant many-body correlations have the general form  $C=\langle \prod^{N_s}_i \sigma^z_i \rangle$, such that the $N_s$-spin product is equal to a product of $N_p$ defect variables
\begin{equation}
  \prod^{N_s}_i \sigma^z_i = \prod^{N_p}_j\tau^x_j.
\end{equation}
To construct such products, we define a transformation $\mathcal{T}_i$ whose action on a product of spin operators is to replace the operator $\sigma_i$ by $\sigma_j \sigma_k$, where $ijk$ is a triangular plaquette hosting a $3$-spin interaction. A general $N$-spin correlation function  of the form 
\begin{equation}
C(r)=\langle \sigma^z_i \prod_j \sigma^z_j \rangle =\langle \sigma_i^z (\prod_j\mathcal{T}_j) \sigma_i^z \rangle
\label{eq:manybody_corr}
\end{equation}
for an arbitrary sequence of transformations given by $\prod_j \mathcal{T}_j$ is always equal to a product of defect variables and can be used to diagnose fractal symmetry breaking.

The behaviour of these correlation functions have a passing resemblance to Wilson loops in lattice gauge theories, where perimeter and area laws diagnose confined and deconfined phases~\cite{YizhiEA}. 
In the fractal-symmetry-broken phase $C\sim \exp(-a N_s)$ where $a>0$ is an $O(1)$ constant. In the symmetric phase such correlations have an `area-law' decay $C\sim \exp(-bN_p)$, where $b$ is another constant. Often, one can construct correlations where $N_s\sim O(1)$ (motivating the name `corner-law') though $N_p$ is macroscopic, making such correlation functions a useful diagnostic for fractal symmetry breaking. In the rest of the section we describe the construction of such correlation functions with $N_s \sim O(1)$.

We  present the construction  for trillium in detail, and briefly outline how the construction for HHK  proceeds along similar lines.
For trillium, we propose a many-spin correlation function between $\delta$ and $\gamma$ spins (chosen for specificity, though we could choose sublattices from any two consecutive slices in Eq.~\eqref{eq:levels})
\begin{equation}
C(r)=\langle \sigma^z_{\delta}(0,0,0) \prod_{ij}\sigma^z_{\delta}(i,j,r) \prod_{ij}\sigma^z_{\gamma}(i,j,r)\rangle.
\label{eq:corr}
\end{equation} 
The products of $\delta$ and $\gamma$-spins appearing in $C(r)$ are determined by: {(i)} defining $\mathcal{T}_i$ (Eq.~\eqref{eq:manybody_corr}) to act on a spin $\sigma^z_i$ to give $\sigma^z_j$ and $\sigma^z_k$, such that $ijk$ is a triangular plaquette and $jk$ come from the two slices lying next to $i$ in the scheme of Eq.~\ref{eq:levels}; and {(ii)} requiring that one starts with $\sigma^z_{\delta}(0,0,0)$ and keeps applying the transformations $\mathcal{T}_i$ until all resulting spins are $\delta$ and $\gamma$ spins in the layer $z=r$. In fact, the terms in such product can be described by the polynomial $M^{-r} (1,0)^{T}$, where $M$ is the transition matrix in Eq.~\ref{eq:CA}.  For matrix CA, these products have a macroscopic number of terms and the resulting $C(r)$ does not reduce to a few-spin correlation function even for special values of $r$. To circumvent this, we construct another correlation function
\begin{equation}
C'(r)=\langle \sigma^z_{\delta}(0,0,0) \prod_{ij}\sigma^z_{\delta}(i,j,-r) \prod_{ij}\sigma^z_{\gamma}(i,j,-r)\rangle.
\label{eq:corr_prime}
\end{equation} 
Here, the $\delta$ and $\gamma$ spins appearing in the product are determined by choosing the transformation $\mathcal{T}_i$ such that they replace a spin $\sigma^z_i$ by $\sigma^z_j \sigma^z_k$, such that $ijk$ is a triangular plaquette and $jk$ come from the two slices preceding  the slice which $i$ belongs to. These transformations are also applied until all the spins in the resulting product involve $\delta,\gamma$ spins at $z=0$. This product is described by the polynomial $M^r (1,0)^T$.  Like $C(r)$, $C'(r)$ also contains a macroscopic number of spins. However, as we saw in Sec.~\ref{subsec:trillium_fractons}, $M^r+M^{-r}$ has $O(1)$ terms for all $r=2^n$. Consequently, we can construct a few-spin correlation function by taking the product of $C(r)$ and $C'(r)$, shifted in the $z$-direction by $\mp r$
\begin{equation}
C_f(r)=C(r)C'(r)=\langle \sigma^z_{\delta}(0,0,-r) \sigma^z_{\delta}(0,0,r) \prod_{ij}\sigma^z_{\delta}(i,j,0)\rangle.
\label{eq:corr_few}
\end{equation}
The $xy$-coordinates of the spins appearing in the product are described by the polynomial $(M^{r}+M^{-r}) (1,0)^T$.
This is reminiscent of the construction of fractonic elementary excitations---for $r=2^n$, we now have a 7-spin correlation functions of $\delta$-spins lying at unit cell positions  described by the corners and center of a regular octahedron. Similarly, we have $7$-spin correlation functions for spins on other sublattices. 

For HHK, we can similarly define many-spin correlation functions which reduce to a correlation function of O(1) spins for $r=2^n$
\begin{align}
\nonumber  C_f(r)={}&\langle \prod_{\{i',j',k'\}}\Big(\sigma^z_{\alpha^{k'}}(i',j',-r) \sigma^z_{\alpha^{k'}}(i',j',r) \Big)\\
        & \times \prod_{ijk}\sigma^z_{\delta}(i,j,0)\rangle.
\label{eq:hhk_corr_few}
\end{align}
Here, the labels ${i',j',k'}$ denote a  group of $xy$-coordinates and sublattices of spins which appear in the product at coordinates $z=-r$ and $z=r$. All labels $k'$ denote sublattices from the same layer $L_a$(according to the notion of layers introduced in Eq.~\eqref{eq:levels_hhk}).  The $xy$-coordinates and sublattices of spins in the product $\prod_{ijk}\sigma^z_{\delta}(i,j,0)$ are described by the polynomials $(M^r_a+M^{-r}_a)\phi_a$, where the vector $\phi_a$ has polynomials over $\ff$ which denote the $xy$ coordinates and sublattices of the spin-product $\prod_{\{i',j',k'\}}\sigma^z_{\alpha^{k'}}(i',j',-r)$. 
For $r=2^n$, if we choose the vector $\phi_a$ describing the spin-product to be ones chosen as initial conditions used to construct EFCs in Tab.~\ref{tab:hhk_fractons}, $C_f(r)$ becomes an 8-spin product, with the spins lying at the corners of a regular octahedron, with vertices at $(\pm \tn,0,0), (0,\pm \tn,0)$ and $(0,0,\pm \tn)$. Thus, we have constructed few-spin  correlation functions (7-spin correlations for trillium and  8-spin  correlations for HHK), which serve as clean diagnostics for fractal symmetry breaking.

\subsection{Absence of fracton topological order under  $F$-$S$ duality}
We briefly discuss why $F$-$S$ duality cannot be leveraged to construct fracton models from the triangular plaquette models considered here. The $F$-$S$ duality is a framework through which commuting Hamiltonians $H_{\text{fracton}}$ with fracton topological order can be obtained from models where  classical spin Hamiltonians $H_{\text{classical}} = \sum_{i,a} \mathcal{O}_{i,a}[\sigma^z]$ with subsystem symmetries are perturbed with a transverse field~\cite{Vijay_Haah_Fu}. This procedure, which is closely related to the duality construction in Sec.~\ref{subsec:crystal_trillium},  involves mapping each interaction term $\mathcal{O}_{i,a}[\sigma^z]$ in $H_{\text{classical}}$ to a {\it nexus spin} $\tau^z_{i,a}$, where $i,a$ are unit cell and sublattice indices. The $A$-terms in $H_{\text{fracton}}$ are {\it nexus charge} operators, and represent the $F$-$S$ dual of the transverse field term $\sigma^x_{i,a}$. Explicitly, each $A_{i,a}[\tau^x]$ consists of the product of $\tau^x$ that are $F$-$S$ dual to interaction terms $\mathcal{O}[\sigma^z]$ that anticommute with $\sigma^x_{i,a}$. Under this procedure, the NM, trillium and HHK models all have nexus charges that themselves form a triangular plaquette network (with each nexus spin participating in three nexus charges). In fact, NM and trillium are self-dual. The obstruction to building $H_{\text{fracton}}$ lies in finding a suitable ``$B$-term'', i.e. a local product of $\sigma^z$ that commutes with the nexus charges. This is impossible since we know that the triangular plaquette models (on the dual side) have fractal subsystem symmetries but no gauge symmetries. More straightforwardly, the ineffectiveness of $F$-$S$ duality stems from the fact that no local product of interaction terms in the classical spin system is the identity.

\section{Concluding Remarks}

We have introduced simple models of Ising spins with three-spin Baxter-Wu type interactions,  that host classical fractal spin liquid states on two lattices of corner sharing triangles: trillium and hyperhyperkagome (HHK). These models are described by fractal symmetries generated by a subset of spin flips prescribed by the action of a matrix CA.  Accordingly, they host (for nearly all system sizes) a macroscopic but subextensive ground state degeneracy, that may be rationalized in terms of a set of spin flips relative to the uniform configuration  generated by applying the matrix CA to an initial two-dimensional pattern of flipped spins  relative to it.
   We present an exact solution of the thermodynamics of these models by  mapping the problem to a gas of defects; in both cases, the thermodynamics are trivial, i.e. the partition function is essentially that of the free defect gas. We also construct low energy excitations which are clusters of  immobile ``fracton" defects. Such elementary fracton clusters (EFC) are created by flipping a fractal subset of spins in the ground state, which are generated by the action of the same matrix CA composed with a finite number of additional spin flips. We present an analytical calculation of the fractal dimension of the fractal subsets of spins flipped in an EFC. We also show that creating (or annealing away) such EFCs involves activation barriers which scale logarithmically with their size, leading in turn  to glassy behaviour under classical single spin-flip dynamics, as verified by classical Monte Carlo simulations. We also construct many-body correlation functions which serve as a diagnostic for spontaneous fractal symmetry breaking, which we anticipate will be a feature of these models in the presence of quantum fluctuations introduced by means of a transverse field. For trillium, we show that a self-duality of the lattice pins such a putative phase transition to a fixed value of the transverse field.

   While many of these conclusions are familiar from the two-dimensional Newman-Moore (NM) model (which has the same three-spin  interaction) and its three-dimensional generalisations (which generically involve higher-spin interactions), we emphasize that the models considered here represent a distinct type of classical fractal liquid. Ultimately, the distinction stems from the fact that the ground states of our models are described by CAs with a matrix transition function acting on a state vector which encodes spin configurations of different sublattices. Unlike the NM model and its generalizations, this  CA is invertible. Most of the conclusions in the NM model rely on a simplification called the ``Freshman's dream" that relies in an essential way on the fact that the  CA transition function is a polynomial over $\ff$. In our case, the CA describing the ground state is a \textit{matrix} of polynomials over $\ff$,  and Freshman's dream is immediately lost. We instead find an alternative procedure to construct simplifying fractal subsets by exploiting the Cayley-Hamilton theorem to construct EFCs. This involves gluing together a CA evolution for $\tn$ steps from a single defect with an inverse-CA evolution from another defect which is $2^{n+1}$ steps away, such that the gluing together creates an O(1) number of defects  independent of the length of the CA evolution. We use related techniques to analytically compute the associated fractal dimension, and similarly to derive the other results summarized above.

Such matrix CAs have been previously investigated in the context of Clifford Quantum CA~\cite{Gutschow,GutschowEA}. In this language, the configuration space of the CA  acting on a two-component vector is identified with a string of Pauli operators.  The standard route to constructing Hamiltonians with ground states described by such matrix CA, as presented in Refs.~\cite{TrithepEA,Yoshida}, generically leads to the presence of  multi-spin interactions of high order. For example, we would need both three- and four- spin interactions for the CA describing ground states in trillium,  and  five-spin interactions for that describing  HHK. Interestingly, our Hamiltonians have a sublattice structure, which result in purely three spin interactions leading to ground states described by such matrix CA. This simplification is significant in light of recent proposals to realize NM Hamiltonians experimentally~\cite{MyersonJainEA}.
While  three-spin interactions  necessarily involve the breaking of time-reversal symmetry to be realized in a magnetic system,  the authors have argued that similar NM-type interactions can be engineered in near-term experiments on Rydberg atom arrays. The proposal involves  placing a set of auxiliary atoms at the centre of each triangular plaquette of ``target" atoms, and engineering repulsive interactions between the atoms such that the low energy subspace can be described in terms of fractal symmetries. In contrast to the more intricate engineering required to extend this proposal to three dimensions via existing 3D fractal models such as NM, the three-spin trillium and HHK models present a case where such an extension is straightforward as long as the atoms can be  trapped in the appropriate 3D lattice.

Our work suggests a few natural directions for the immediate future. It would be interesting to explore quantum phase transitions involving spontaneous breaking of fractal symmetries described by matrix CA, and this will hopefully shed some light on the critical theory of such transitions, as raised by the investigations of Ref.~\cite{YizhiEA}.  It would also be interesting to design stabilizer Hamiltonians which host topological phases  with our ``matrix fractons" as their low energy excitations. With the  optimism that is appropriate to a conclusions section, we defer these to future work.

\begin{acknowledgements}
We  thank J.~Reuther, J.P.~Garrahan, M.~Fava,  T.~Devakul and Y.~You  for insightful discussions. We are especially grateful to J.T. Chalker for bringing Ref.~\cite{ChillalEA} to our attention, sparking this line of inquiry. We acknowledge support from  the European Research Council under
the European Union Horizon 2020 Research and Innovation Programme via Grant Agreement No. 804213-TMCS.
\end{acknowledgements}

\appendix
\section{Fractal dimensions}
\label{app:fracdim}
In this appendix, we present a calculation of the fractal dimensions of the sets of spins which are flipped to create the EFCs presented in Sec.~\ref{sec:fractons}. These calculations are feasible due to simplifications afforded by Eq.~\eqref{eq:trace_eq} (for trillium), and Eqs.~\eqref{eq:hhk_traceq} and \eqref{eq:hhk_traceq12} (for HHK), and generalizes the  approach of Ref.~\cite{Gutschow_Nesme_Werner}. As before, we first present the calculation for trillium, and then  outline the calculation for HHK with an emphasis on differences between the two cases.

Interpreting the direction of evolution of CA as `time', we use the term `spacetime diagram' for the patterns of down spins generated by a CA. For trillium, patterns of down spins in ground states and associated defects are described by a matrix CA with transition matrix $M$ (Eq.~\ref{eq:CA}). $M$ satisfies Eq.~\eqref{eq:trace_eq} (repeated for convenience)
\begin{align}
  \tag{\ref{eq:trace_eq}}
\nonumber M^{2^n}&=M^{-2^n}+\Tr(M)^{2^n}\\
\nonumber \Tr(M)&= 1+x+\xb +y +\yb.
\end{align}

We denote the spacetime diagram generated by the action of $\tn$ steps of the CA on an initial configuration $\psi_0$ at $z=0$ by $\dm (\tn)$. The $\psi_0$ dependence has been suppressed because of its irrelevance to the fractal structure for large values of $n$. 
From $z=0$ until $z=\tn$ the diagram $\dm(2^{n+1})$ is given by $\dm(\tn)$. We will now use Eq.~\eqref{eq:trace_eq} to express the rest of $\dm(2^{n+1})$ in terms of superpostions of $\dm(\tn)$.
Eq.~\eqref{eq:trace_eq} implies that action of $M^{2^n}$ on an arbitrary initial configuration $\psi_0$ at $z=0$  leads to a configuration at $z=\tn$ which can be described as the superposition of two terms:
\begin{enumerate}
  \item $M^{-2^n} \psi_0$, which would evolve back to $\psi_0$ under a further evolution by $2^n$.
  \item $(1+x^{2^n} +y^{2^n} +\xb^{\tn}+\yb^{\tn})\psi_0 $ which describes the superposition of five copies of the initial configuration, shifted by $(0,0),(0,\pm\tn),(\pm \tn,0)$.
  \end{enumerate}

To find a recursion relation for $\dm (2^{n+1})$ in terms of  $\dm (\tn)$, we now consider the result of CA evolution by $\tn$ further steps on each of the two terms separately:
\begin{enumerate}
  \item The first term $M^{-\tn} \psi_0$ evolves back to $\psi_0$ under the action of $M^{\tn}$. We denote the spacetime diagram thus generated as $\dmb(\tn)$. 
  \item The evolution of the second term, $(1+x^{2^n} +y^{2^n} +\xb^{\tn}+\yb^{\tn})\psi_0$, describes 5 copies of $\dm$, starting at $z=\tn$ and shifted by coordinates $(0,0),(0,\pm\tn),(\pm \tn,0)$ in the $xy$ direction. 
\end{enumerate}

\begin{figure}[t]
    \includegraphics[width=\textwidth]{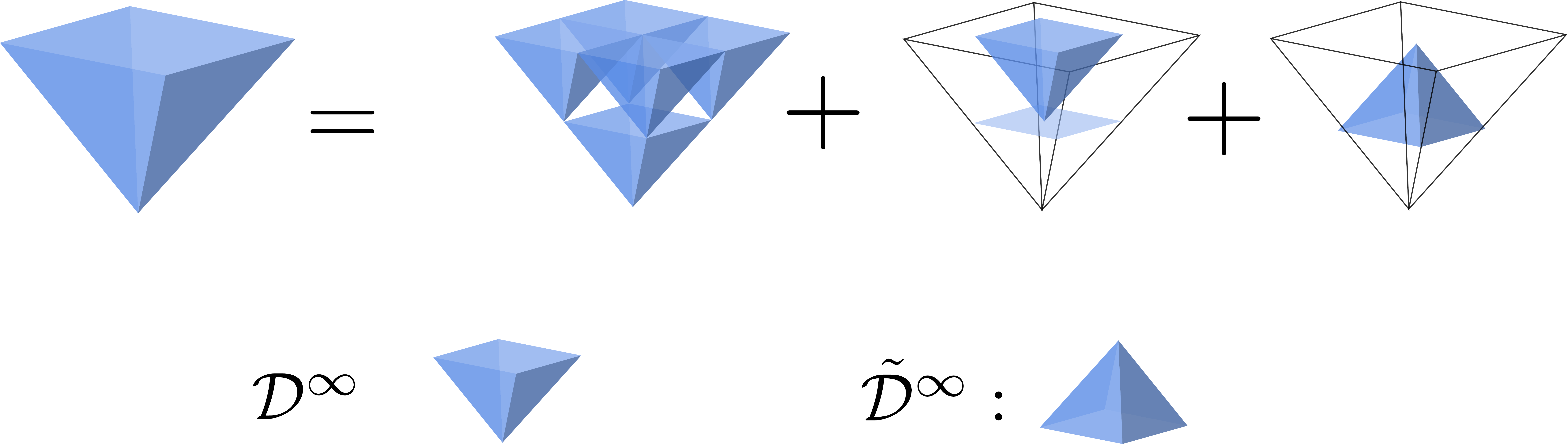}
    \caption{An inflation rule for $D^{\infty}$, the spacetime diagram generated in the long-time evolution of the CA described by the transition matrix $M$ (Eq.~\eqref{eq:CA}) relevant to trillium. $D^{\infty}$ is expressed as a superposition ($\text{mod } 2$) of 5 copies of $D^{\infty}$ which are smaller by a factor  of 2, and 1 inverted copy, denoted by $\dmb^{\infty}$, which is also smaller by a factor of 2. }
    \label{fig:pyramid_recursion}
\end{figure}

\begin{figure}[t]
    \includegraphics[width=\textwidth]{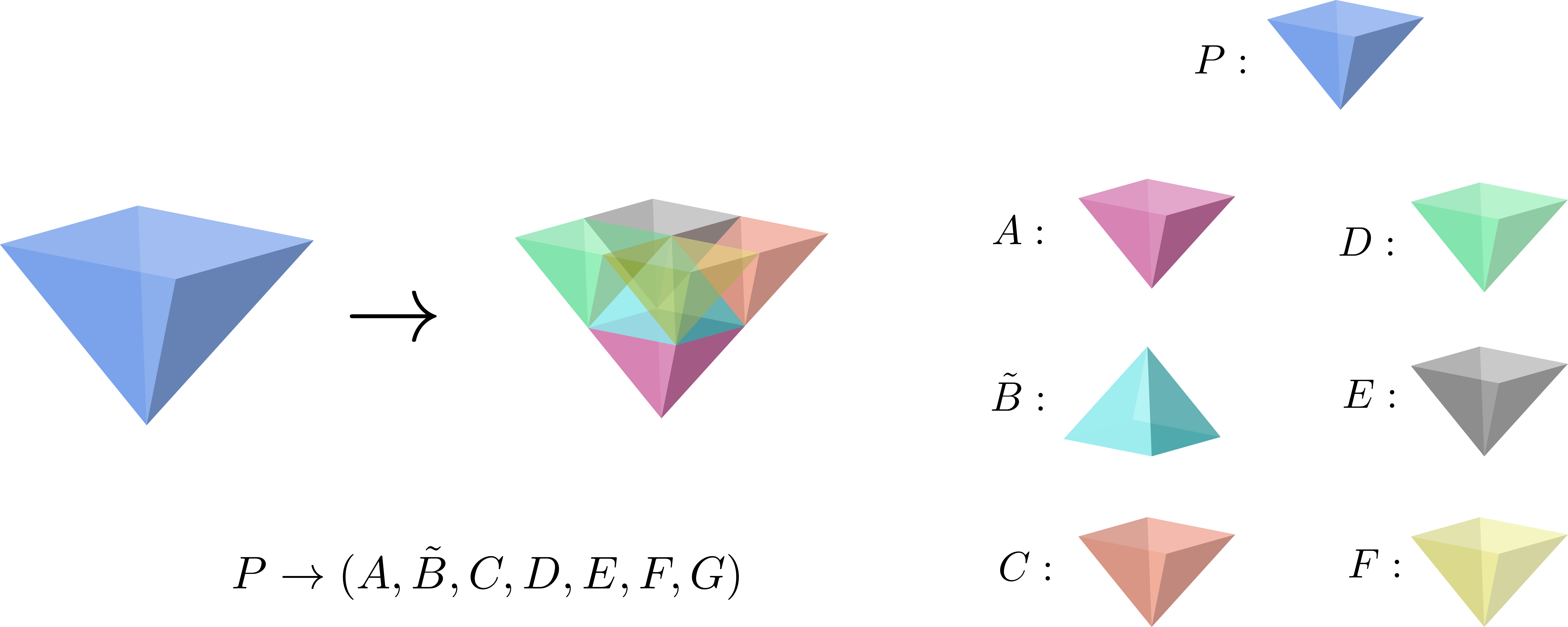}
    \caption{ The decomposition of a pyramid of height 2 into six non-overlapping pyramids of height 1. This decomposition is convenient for inflation rules, and is the pictorial representation of Eq.~\ref{eq:schematic_inflation_rule}.  }
    \label{fig:pyramid_decomposition}
\end{figure}

Therefore, $\dm(2^{n+1})$ is a superposition of 6 copies of $\dm(\tn)$ (5 from the evolution of the second term, as well as one from the  initial evolution until $z=\tn$) and one copy of $\dmb(\tn)$ (from the evolution of the first term). We now define $\dm^{\infty}$ to be the shape obtained from the large-$n$ limit of $\dm(\tn)$ rescaled by $\tn$. The form of  $\Tr(M)$ suggests that such a limit shape is a square-based pyramid of height $1$, with vertices at $(0,0,0), (0,\pm 1,1), (\pm 1, 0, 1)$. The recursion relation expressing $\dm(2^{n+1})$ in terms of superposition of $\dm (2^{n})$  and $\dmb(\tn)$ becomes an \emph{inflation rule}: a relation expressing a copy of $\dm^{\infty}$ inflated by a factor of $2$ as a superposition of 6 copies of itself and one copy of $\dmb^{\infty}$. In anticipation of the usefulness of this rule for the rest of the section, we call this the ``master inflation rule". The master inflation rule is convenient to express diagramatically, and we display it in Fig.~\ref{fig:pyramid_recursion}. To express this, we introduce the symbol
 $\dm^{\infty}_{k}(x,y,z)$ to denote the limit-pyramid inflated by a factor of $k$, with the apex at $(x,y,z)$ and the vertices of the square base at $(x\pm k,y,z+k), (x,y\pm k,z+k)$. Similarly, we use $\dmb^{\infty}(x,y,z)$ for the inverse evolution. In terms of this, the master inflation rule expresses the limit-pyramid as:
 \begin{align}
   \nonumber \dm^{\infty}_{2}(0,0,0) ={}& \dm^{\infty}_{1}(0,0,0)+\dm^{\infty}_{1}(0,0,1)\\
\nonumber & +\dm^{\infty}_{1}(1,0,1) +\dm^{\infty}_{1}(-1,0,1)\\
\nonumber & +\dm^{\infty}_{1}(0,1,1)+\dm^{\infty}_{1}(0,-1,1) \\
&+\dmb^{\infty}_{1}(0,0,2).
\label{eq:pyramid_recursion}
 \end{align}
 Working directly with master inflation rule is problematic because the multiple copies of $\dm^{\infty}$ and $\dmb^{\infty}$ interfere with each other in a non-trivial way. We want to express a pyramid inflation rule for a general pyramidal shape $P_2$ of height $2$ in terms of 6 \emph{non-overlapping pyramids} of height $1$ as follows\footnote{Note that 6 pyramids of height 1 do not fill up the space in a pyramid of height 2. However, for the shapes described by $\dm^{\infty}$, this description is sufficient.}:
 \begin{align}
\nonumber P_{2}(0,0,0) &= A_{1}(0,0,0)\\
\nonumber & +C_{1}(1,0,1) +D_{1}(-1,0,1) \\
\nonumber &+E_{1}(0,1,1)+F_{1}(0,-1,1)\\
&+\tilde{B}_{1}(0,0,2).  
\label{eq:schematic_inflation_rule}
 \end{align}
 The arguments denote coordinates of the apex of the pyramid, and symbols with a tilde denote inverted pyramidal limit-shapes. This decomposition of a pyramid of height $2$ in terms of 6 non-overlapping pyramids is shown in Fig.~\ref{fig:pyramid_decomposition}.
 We will also the following shorthand for Eq.~\eqref{eq:schematic_inflation_rule}:
 \begin{equation}
   P \rightarrow (A,\tilde{B},C,D,E,F,G).
 \end{equation}
 We now express the master inflation rule (Eq.~\eqref{eq:pyramid_recursion}, Fig. ~\ref{fig:pyramid_recursion}) as  $\dm^{\infty}\rightarrow (A,\tilde{B},C,D,E,F)$. The price for the expression in terms of non-overlapping pyramids is the complication that the pyramids $A,\ldots,F$ are are not as simply related to $\dm^{\infty}$.
However it is clear that the shape $A$ is same as $\dm^{\infty}$. This gives us:
 \begin{equation}
 A \rightarrow (A,\tilde{B},C,D,E,F).
 \label{eq:inflation_A}
 \end{equation}
 To specify how the pyramid $A$ undergoes mutliple inflations, we must specify how the pyramids appearing on the RHS of Eq.~\eqref{eq:inflation_A} inflate. Before we carry this forward, we note that
 like $M$ in Eq.~\ref{eq:trace_eq}, $M^{-1}$ also follows $M^{-\tn}=\Tr(M)^{\tn}+M^{\tn}$. Therefore $\dmb^{\infty}$, the limit-pyramid of spacetime diagrams generated by $M^{-1}$, has a similar master inflation rule, and can be decomposed similarly. We use the symbol $\tilde{A}$ for $\dmb^{\infty}$,  and write its inflation rule as 
 \begin{equation}
\dmb^{\infty}= \tilde{A} \rightarrow (\tilde{A},B,\tilde{C},\tilde{D},\tilde{E},\tilde{F}).
 \label{eq:inflation_Ain}
 \end{equation}
 We emphasize that the limit-pyramids $A,B,\ldots, F$ are different from $\tilde{A},\tilde{B}, \ldots ,\tilde{F}$. We now use the inflation rules for $\dm^{\infty}$(Eq.~\eqref{eq:inflation_A}) and $\dmb^{\infty}$(Eq.~\eqref{eq:inflation_Ain}) in the master inflation rule of Eq.~\eqref{eq:pyramid_recursion} (Fig.~\ref{fig:pyramid_recursion}) to have inflation rules for all the other pyramids $B\cdots F$

 \begin{align}
   \nonumber \tilde{B} &\rightarrow (\tilde{A}+\tilde{B},A+B,\tilde{C},\tilde{D},\tilde{E},\tilde{F})\\
 \nonumber C &\rightarrow (A,\tilde{B},C,C+D,E,F)\\
 \nonumber D &\rightarrow (A,\tilde{B},C+D,D,E,F)\\
 \nonumber E &\rightarrow (A,\tilde{B},C,D,E+F,F)\\
           F &\rightarrow (A,\tilde{B},C,D,E,E+F).
           \label{eq:inflation_rules_1}
 \end{align}
 Each inflation rule above has a counterpart obtained by interchanging  all symbols and their versions with a tilde.
 Inflation rules for superpositions like $C+D$ can be found by superposing the rules for $C$ and $D$
 \begin{align}\label{eq:inflation_rules_2}
 	\nonumber C    &\rightarrow (A,\tilde{B},C,C+D,E,F)\\
 	\nonumber D    &\rightarrow (A,\tilde{B},C+D,D,E,F)\\
 	\nonumber C+ D &\rightarrow (0,0,D,C,0,0)\\
 	\nonumber E+ F &\rightarrow (0,0,0,0,F,E)\\
 	A+ B &\rightarrow (B,\tilde{A},0,0,0,0).
\end{align}
 As before, each inflation rule has a counterpart obtained by interchanging each symbol and its counterpart with a tilde. Therefore, we have a set of inflation rules which map a set of 18 symbols, $(A,B,C,D,E,F,C+D,E+F,A+B)$ and their counterparts with a tilde, to themselves.
 Starting from a diagram $A$, one can define a counting vector $\zeta_0=(N_A=1,0,0,\ldots)$ which counts the number of occurrences of each type of symbol, each of them corresponding to a particular limit-pyramid. The inflation rules (Eq.~\eqref{eq:inflation_A},\eqref{eq:inflation_rules_1} and \eqref{eq:inflation_rules_2}) determine a \emph{substitution matrix} $S$, such that  applying the inflation rules $n$ times gives us the counting vector.
 \begin{equation}
   \zeta_n= S^n \zeta_0.
 \end{equation}
In the limit of large $n$, the leading term of the counting-vector is given by  $\lambda_0^n U_0$, where $\lambda_0$ and $U_0$ are the leading right eigenvalue and right-eigenvector of $S$. The volume of the resulting spacetime diagram is given by $\lambda_0^n V^T U_0$, where $V$ is a vector, as yet not determined, with the volume of each of the 18 symbols as its elements. 
The fractal dimension of this diagram is therefore given by $d_f=\log_2(\lambda_0)$. In our case, we do not need to distinguish between a symbol and its `tilde'-ed counterpart in the counting vector. This results in a $9\times9$ matrix, whose eigenvalues are $(-2,-2,0,1,1,1,1,)$ and $(5 \pm \sqrt{33})/2$. Therefore the fractal dimension of the spacetime diagrams generated by our CA for trillium is given by 
\begin{equation}
  d_f=\log_2\Big[\frac{5+\sqrt{33}}{2}\Big] \sim 2.425.
\end{equation}

\begin{table}[t]
  \centering
      \begin{tabular}{|c|c|}
        \hline 
	
	Initial condition  & Limit-pyramid generated by CA  \\ 
	\hline \hline
	$(1,0,0,0)$ & $A$ \\ 
	\hline
	$(0,1,0,0)$ & $B$ \\
	\hline
	$(0,0,1,0)$ & $C$ \\
	\hline
	$(0,0,0,1)$ & $D$ \\
	\hline
	$(0,1,1,0)$ & $E$ \\
	\hline
	$(0,1,0,1)$ & $F$ \\
	\hline
	$(xy,1,1,0)$ & $G$ \\
	\hline
	$(xy,1,0,1)$ & $H$ \\
	\hline
	$(xy,1,1,1)$ & $I$\\
	\hline
	\end{tabular}
	\caption{Initial conditions and the symbols for limit-pyramids generated by the long-time CA evolution starting from them. These symbols are used to express the inflation rules of Eq.~\eqref{eq:hhk_inflationrules}}
\label{tab:limit_shapes_hhk}
\end{table}

The corresponding calculation for HHK can be carried out similarly. For specificity we consider the spins in layer $L_0$ (Eq.~\eqref{eq:levels_hhk}), which consists of spins in sublattices $\alpha^0, \alpha^5, \alpha^8$ and  $\alpha^9$.  The CA evolution of spins in this layer is described by the matrix $M_0$, which satisfies Eq.~\eqref{eq:hhk_traceq} (repeated for convenience)
\begin{align}
\nonumber  M_0^{\tn}+M_0^{-\tn}=
  \tag{\ref{eq:hhk_traceq}}
\begin{pmatrix}
  f^{\tn} & 0 & xy T^{\tn} & 0 \\
  \xb \yb T^{\tn} & f^{\tn} & T^{\tn} & T^{\tn} \\
  \xb \yb T^{\tn} & 0&  g^{\tn}  & 0 \\
   0 & T^{\tn}&  T^{\tn} & g^{\tn} \\
 \end{pmatrix} ,\\
\nonumber  \text{where }, f=x+\xb, g=y+\yb, \text{when } n \text{ is odd},\\
\nonumber            f=y+\yb, g=x+\xb, \text{when } n \text{ is even}. \\
\nonumber T= \Tr(M_0)= 1+x+\xb +y +\yb.
\end{align}
Here, we will need to distinguish between the spacetime diagrams generated by different initial conditions, though they will turn out to have the same fractal dimensions. 
As before, we will work with limit-shapes obtianed by taking the large-$n$ limit of spacetime diagrams obtained by CA evolution of $\tn$ steps, and rescaling them by $\tn$. Let us denote the limit-shapes obtained from initial conditions $(1,0,0,0)^T$, $(0,1,0,0)^T$, $(0,0,1,0)^T$ and $(0,0,0,1)^T$ by $A_1$, $B_1$, $C_1$ and $D_1$ respectively. $A_k$ denotes the limit-shape $A_1$ inflated by a factor of $k$. We will also use symbols like $\tilde{A}_k$ to denote the corresponding limit-shapes generated, when the inverse CA with transition matrix $M_0^{-1}$ acts on initial conditions which generate $A_k$ .

Now, imagine an initial condition $\phi^0_0=(0,1,0,0)$, which describes a single down spin in the $\alpha^5$ sublattice at unit cell coordinates $(0,0,0)$ and generates the limit shape $B$ in the large-$n$ limit. If we take $n$ to be odd, starting with $\phi^0_0$, the flipped spins at $z=2^n$ after CA evolution of $\tn$ steps is given by a superposition of the following terms

\begin{itemize}
  \item $M_0^{-\tn} \phi^0_0$, which  comprises of a macroscopic number of flipeped spins, but evolves back to $\phi^0_0$ under a further evolution of $\tn$ steps.
  \item The vector $(0, f^{\tn},0,T^{\tn})$ describing two flipped $\alpha^5$-spins separated by $\tn$ lattice spacings, and 4 flipped $\alpha^9$-spins  described by the term $T^{\tn}=x^{\tn}+y^{\tn}+\xb^{\tn}+\yb^{\tn}$.
\end{itemize}
The spacetime diagram for $2^{n+1}$ steps can be expressed in terms of spacetime diagrams for $\tn$ steps by considering a further evolution by $\tn$ steps. The first term evolves back to $\phi^0_0$, while the evolution of the second term generates 2 copies of $B_{\tn}$ (from the CA evolution of two flipped $\alpha^5$ spins at $z=\tn$) and 4 copies of $D_{\tn}$ (from the CA evolution of the 4  flipped $\alpha^9$ spins at $z=\tn$).

It is clear from the form of $T=\Tr(M_0)$ that the limit-shapes are again square-based pyramids, with vertices at $(0,0,0)$, $(\pm 1,0,1)$,$(0,\pm 1,1)$. If we focus on odd values of $n$ for the moment, the inflation rules in terms of these limit pyramids are given by
\begin{align}
  \nonumber B_2(0,0,0)={}&B_1(0,0,0)+ \tilde{B_1}(0,0,2)+B_1(-1,0,1)\\
  \nonumber            &+ B_1(1,0,1)+C_1(1,0,0)+C_1(-1,0,0)\\
	      &+C_1(0,1,0)+C_1(0,-1,0).
\end{align}
The arguments denote the coordinates of the apex of the limit-pyramids.  We now introduce a shorthand for this equation, which organizes  inflation rules in terms of different limit pyramids and their apex coordinates.
\begin{align}
  B \rightarrow(B,\tilde{B},B+D,B+D,D,D).
  \label{eq:hhk_inflation_demo}
\end{align}

Here, the limit-pyramid $B+D$  corresponds to a superposition of two limit-pyramids $B$ and $D$ with both their apices at $(\pm1,0,1)$. This corresponds to CA evolution with an initial condition $(0,1,1,0)^T$. 
To specify how any limit-pyramid undergoes multiple inflations, we must find out how the other limit-pyramids appearing in Eq.~\ref{eq:hhk_inflation_demo}, including $B+D$, inflate. One must also find out inflation rules for any new limit-pyramids which appear in the process. 
To express these inflation rules concisely, we introduce new symbols $E,\ldots,I$ for these limit shapes which result out of superposition of the shapes $A,\ldots,D$, or equivalently, arising out of CA evolution from superposition of different initial conditions which generate $A,\ldots,D$ . We tabulate these initial conditions along with symbols for the generated limit-pyramids in Tab.~\ref{tab:limit_shapes_hhk}.
For different limit pyramids, we can obtain the following inflation rules:
\begin{align}
\nonumber  A &\rightarrow (A,\tilde{A}, G,  G, E, E)\\ 
\nonumber  B &\rightarrow (B, \tilde{B},F,F,D,D)\\
\nonumber  C &\rightarrow (C,\tilde{C},H,H,I,I) \\
\nonumber  D &\rightarrow (D, \tilde{D},B,B,F,F)\\
\nonumber  E &\rightarrow (E, \tilde{E},A,A,G,G)\\
\nonumber  F &\rightarrow (F, \tilde{F},D,D,B,B)\\
\nonumber  G &\rightarrow (G, \tilde{G},E,E,A,A)\\
\nonumber  H &\rightarrow (H, \tilde{H},I,I,C,C)\\
I &\rightarrow (I, \tilde{I},C,C,H,H).
\label{eq:hhk_inflationrules}
\end{align}
We have another set of inflation rules where all limit-pyramids in Eq.~\eqref{eq:hhk_inflationrules} are interchanged with their counterparts with a tilde.

A crucial property of Eq.~\eqref{eq:hhk_inflationrules} is that each limit pyramid inflated by a factor of 2 is composed of six entirely non-overlapping limit-pyramids.  While the inflation rules of Eq.~\eqref{eq:hhk_inflationrules} change for even $n$, and if spins in other layers ($L_1$ and $L_2$) are considered, this fact still continues to hold. This is sufficient to determine the fractal dimension of these limit-pyramids, and consequently, the fractal dimension of the set of spins flipped to create the EFCs for HHK. To see this, note that in the limit-pyramid inflated by $n$ built of $6^n$ non-overlapping uninflated limit-pyramids. Irrespective of any other details of the structure of these limit pyramids, this implies a fractal dimension of 
\begin{equation}
  d_f= \log_2 (6) \sim 2.58.
\end{equation}

\begin{figure}[t]
    \includegraphics[width=\textwidth]{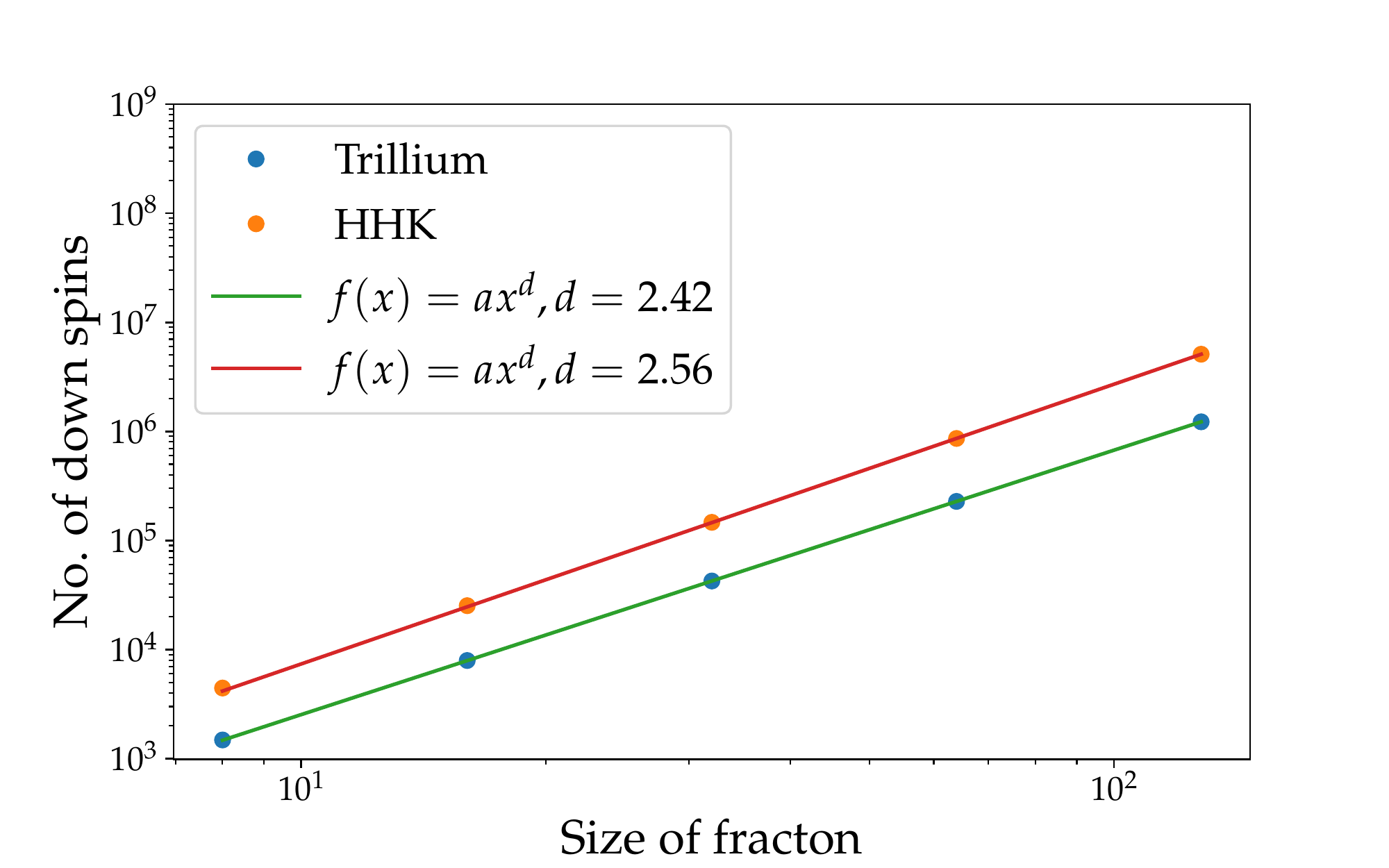}
    \caption{ The total number of down spins in elementary fractonic clusters (EFCs) plotted against the linear size of such particles in a logarithmic scale, for both trillium and HHK. }
    \label{fig:fracdim}
\end{figure}

We have verified  the analytical results  for the fractal dimensions of EFCs of size $2^n$ by comparing against those obtained by explicitly constructing the relevant fractal spin flip patterns, finding excellent agreement as shown in Fig.~\ref{fig:fracdim}.

\section{More on HHK}

In this appendix we supply details of various statements about HHK that were only given telegraphically in the main text.
\subsection{Absence of self-duality}
\label{app:hhk_no_self_duality}
While trillium is self-dual in the sense described in Sec.~\ref{subsec:crystal_trillium}, we demonstrate here that HHK does not satisfy this property. The proof  relies on showing that the connectivities of the real and dual models are different. Consider connecting two direct sites (spins) in HHK with a link whenever they share the same interaction term in the triangular plaquette model. A key feature of HHK is that the smallest non-trivial spin loop is of length 4 (a 4-loop). By inspection, these 4-loops come in triples forming the edges of a triangular prism, where the triangular faces themselves are plaquettes. Let ABCD label the four sequential spins of one of these 4-loops. Spins A and C are connected by two distinct 2-paths (i.e. a path consisting of two links), namely the paths passing through B or D. There are no pairs of direct sites that are connected by more than two distinct 2-paths. Now consider the dual model of HHK, where direct plaquettes are mapped to dual sites, and links are formed between dual sites whenever the corresponding direct plaquettes share a direct spin. Focusing on the dual image of the triangular prism feature mentioned above, we notice that the two dual sites corresponding to the triangular faces of the prism are connected by three distinct 2-paths (via the three dual sites that originate from the three long edges of the prism). Hence the dual lattice model cannot be equivalent to the direct model. 

\subsection{CA transition matrices}
\label{app:hhk_matrices}
As mentioned in Sec.~\ref{sec:ca_hhk}, the CA transition matrices $M_0$, $M_1$ and $M_2$ describe the CA evolution of ground state spin configurations in terms of spin-configurations of sites in $L_0(z)$, $L_1(z)$ and $L_2(z)$ respectively (Eq.~\ref{eq:levels_hhk}, Eq.~\ref{eq:hhk_ca_evol}). The expressions for these matrices and their inverses,  determined from Eqs.~\eqref{eq:hhk_const1}--\eqref{eq:hhk_const3}, are
\begin{widetext}
\begin{align}
M_0(x,y)&=
\begin{pmatrix}
 1 & x+xy & x+xy^{2} & y+xy \\
 0 & x & x+xy & 1+x \\
 \xb\yb+\xb & \xb+1 & \xb+1+y & \xb+\yb \\
 \xb+1 & \xb+\yb+1+x & \xb+\yb+x+xy & \yb \\
\end{pmatrix}\\
  M_0^{-1}(x,y)&=
\begin{pmatrix}
 \xb+1+x & x+xy & y+x^{2}y & y+xy \\
 \xb^{2}\yb+\xb\yb^{2}+\xb+\yb & \xb & \xb+\yb+y+xy & \xb+\yb+1+y \\
 \xb^{2}\yb+\xb\yb^{2}+\xb\yb+\yb & \xb+1 & \xb+\yb+1 & \xb+\yb \\
 \xb+1 & 1+y & y+xy & y \\
\end{pmatrix}\\
  M_1(x,y)&=
\begin{pmatrix}
 x & 1+y & y+x & x+xy+xy^{2}+x^{2}y \\
 \xb\yb+\xb & 1 & \xb+1 & y+xy \\
 \xb\yb+1 & \xb+1 & \xb & 1+y \\
 \xb\yb^{2}+\xb\yb+\xb+\yb & \xb\yb+\yb & \xb\yb+\xb & \yb+1+y \\
\end{pmatrix}\\
  M_1^{-1}(x,y)&=
\begin{pmatrix}
 \xb & 1+y & y+x & y+xy \\
 \xb\yb+\xb & \yb+1+y & \yb+1+y+x & y+xy \\
 \xb\yb+1 & \yb+1+y+x & x & 1+y \\
 \xb^{2}\yb+\xb\yb & \xb\yb+\yb & \xb\yb+\xb & 1 \\
\end{pmatrix} \\
  M_2(x,y)&=
\begin{pmatrix}
 \xb+1+y & \xb\yb+1 & \xb+1+y+x & 1+x \\
 1+x & \yb & 1+x & x+xy \\
 \xb+x & \xb\yb+\yb & \xb+1+x & x+xy \\
 \xb^{2}+\xb\yb+\xb y+\yb & \xb^{2}\yb+\xb\yb^{2}+\xb\yb+\xb & \xb^{2}+\xb\yb+\xb y+1 & \xb \\
\end{pmatrix} \\
  M_2^{-1}(x,y)&=
\begin{pmatrix}
 \xb+\yb+1 & \xb\yb+1 & \yb+1 & 1+x \\
 \xb y+y^{2}+x+xy & y & 1+x & y+xy+xy^{2}+x^{2}y \\
 \yb+y & \xb\yb+\yb & 1 & x+xy \\
 \yb+1 & \xb\yb+\yb & 0 & x \\
\end{pmatrix}
\label{eq:CA_hhk12}
\end{align}
\end{widetext}

\subsection{Non uniqueness of CA description}
\label{app:hhk_partitions}
\begin{figure*}
    \includegraphics[width=\textwidth]{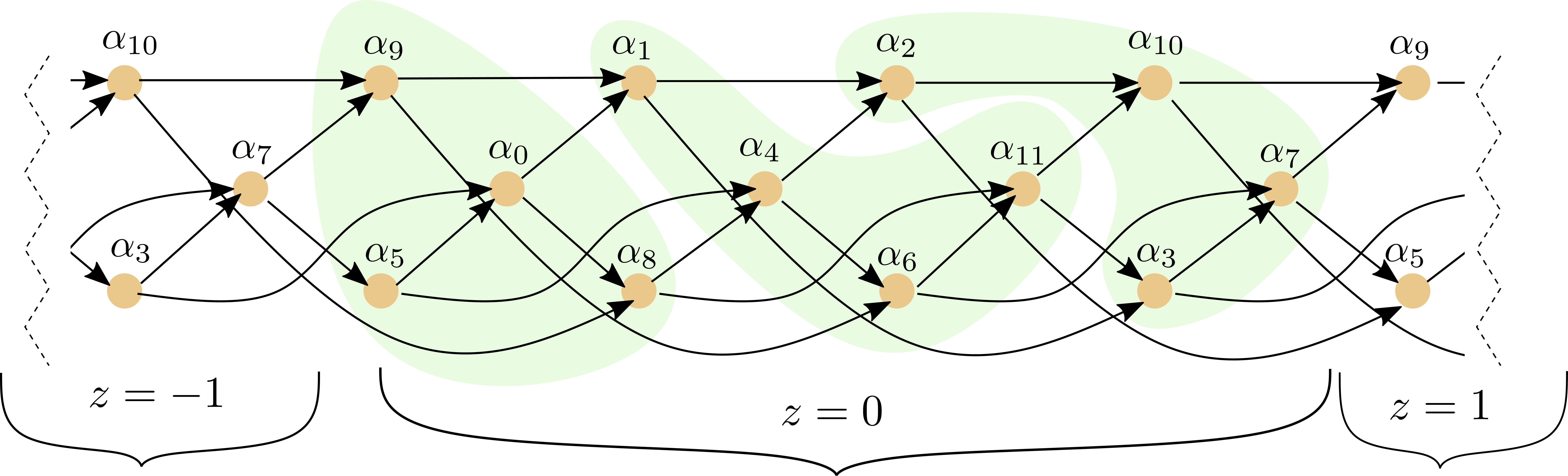}
    \label{fig:hhk_graph}
\end{figure*}
While we describe spin configurations in terms of layers defined in Eq.~\ref{eq:levels_hhk}, the grouping into layers is not unique. To see this let us go back to Eqs.~\eqref{eq:hhk_const1}--\eqref{eq:hhk_const3}. To visualize how spins of a particular sublattice are determined in terms of other sublattices, we associate a directed graph to these equations. Each node of the graph corresponds to a sublattice $\alpha^i$ and the $z$-coordinate. If Eqs.~\eqref{eq:hhk_const1}--\eqref{eq:hhk_const3} determine a spin $s_i$ in terms of spins $s_j$ and $s_k$ by the  relation $s_i=s_j s_k$, such that $s_i,s_j$ and $s_k$ have z-coordinates $z_i,z_j,z_k$ and belong to sublattices $\alpha^i,\alpha^j,\alpha^k$ respectively, then we add a directed edges in the graph from $(\alpha^j,z_j)$ to $(\alpha^i,z_i)$ and from $(\alpha^k,z_k)$ to $(\alpha^i,z_i)$. A section of such a graph is displayed in Fig.~\ref{fig:hhk_graph}. The green shaded regions denote the layers of Eq.~\ref{eq:levels_hhk}. The defining properties of the layers are  that each layer as an ordering of slices, such that each spin is uniquely determined by spins lying the layer preceding it, or in a preceding slice in the same layer as it.
It is evident from the figure that the graph so constructed has a sense of periodicity---the whole graph can be shifted horizontally resulting in a graph with the same connectivities. For \textit{e.g.}, $(\alpha^0,\alpha^1,\alpha^2,\alpha^5,\alpha^4,\alpha^8,\alpha^3,\alpha^7,\alpha^6,\alpha^9,\alpha^{10},\alpha^{11}) \rightarrow (\alpha^4,\alpha^2,\alpha^{10},\alpha^3,\alpha^{11},\alpha^5,\alpha^6,\alpha^0,\alpha^8,\alpha^1,\alpha^{9},\alpha^{7})$ does not change the connectivities. This immediately suggests 4 choices of groupings into layers, including the one presented in Eq.~\eqref{eq:levels_hhk}. However, since the connectivities remaining unchanged, so fo the transition matrices $M_i$ (Eq.~\eqref{eq:hhk_ca_evol}) which describe the CA . It can be verified that descriptions in terms of these different groupings into layers do not give us any new fracton configurations for EFCs beyond those presented in Tab.~\ref{tab:hhk_fractons}.

\subsection{Periodic ground state configurations}
\label{app:hhk_periodic_ground state}
As mentioned before, HHK has 16 periodic spin-configurations which correspond to ground states. This provides us with a lower bound for the number of ground states irrespective of system size and boundary conditions, and rules out the possibility of an one-to-one mapping between spin configurations and defect configurations. Here we list all such periodic ground state configurations (except the trivial ground state where all spins point up), by listing the sublattices which host down spins in each unit cell:
\begin{enumerate}
\item  $\alpha^{3}$,  $\alpha^{5}$,  $\alpha^{7}$,  $\alpha^{8}$,  $\alpha^{10}$,  $\alpha^{11}$ 
\item  $\alpha^{2}$,  $\alpha^{4}$,  $\alpha^{6}$,  $\alpha^{7}$,  $\alpha^{8}$,  $\alpha^{10}$ 
\item  $\alpha^{2}$,  $\alpha^{3}$,  $\alpha^{4}$,  $\alpha^{5}$,  $\alpha^{6}$,  $\alpha^{11}$ 
\item  $\alpha^{1}$,  $\alpha^{4}$,  $\alpha^{8}$,  $\alpha^{9}$,  $\alpha^{10}$,  $\alpha^{11}$ 
\item  $\alpha^{1}$,  $\alpha^{3}$,  $\alpha^{4}$,  $\alpha^{5}$,  $\alpha^{7}$,  $\alpha^{9}$ 
\item  $\alpha^{1}$,  $\alpha^{2}$,  $\alpha^{6}$,  $\alpha^{7}$,  $\alpha^{9}$,  $\alpha^{11}$ 
\item  $\alpha^{1}$,  $\alpha^{2}$,  $\alpha^{3}$,  $\alpha^{5}$,  $\alpha^{6}$,  $\alpha^{8}$,  $\alpha^{9}$  $\alpha^{10}$ 
\item  $\alpha^{0}$,  $\alpha^{5}$,  $\alpha^{6}$,  $\alpha^{7}$,  $\alpha^{8}$,  $\alpha^{9}$ 
\item  $\alpha^{0}$,  $\alpha^{3}$,  $\alpha^{6}$,  $\alpha^{9}$,  $\alpha^{10}$,  $\alpha^{11}$ 
\item  $\alpha^{0}$,  $\alpha^{2}$,  $\alpha^{4}$,  $\alpha^{5}$,  $\alpha^{9}$,  $\alpha^{10}$ 
\item  $\alpha^{0}$,  $\alpha^{2}$,  $\alpha^{3}$,  $\alpha^{4}$,  $\alpha^{7}$,  $\alpha^{8}$,  $\alpha^{9}$  $\alpha^{11}$ 
\item  $\alpha^{0}$,  $\alpha^{1}$,  $\alpha^{4}$,  $\alpha^{5}$,  $\alpha^{6}$,  $\alpha^{7}$,  $\alpha^{10}$,  $\alpha^{11}$ 
\item  $\alpha^{0}$,  $\alpha^{1}$,  $\alpha^{3}$,  $\alpha^{4}$,  $\alpha^{6}$,  $\alpha^{8}$ 
\item  $\alpha^{0}$,  $\alpha^{1}$,  $\alpha^{2}$,  $\alpha^{5}$,  $\alpha^{8}$,  $\alpha^{11}$ 
\item  $\alpha^{0}$,  $\alpha^{1}$,  $\alpha^{2}$,  $\alpha^{3}$,  $\alpha^{7}$,  $\alpha^{10}$ 
  \end{enumerate}

%

\end{document}